\documentclass[prd,aps,twocolumn,preprintnumbers,showpacs, nofootinbib,superscriptaddress,notitlepage]{revtex4-1}
\usepackage{amsmath,graphicx,epsfig,amssymb}
\usepackage{inputenc}
\usepackage[usenames]{color}
\usepackage{slashed}
\usepackage{multirow}
\usepackage{subfigure}
\usepackage{dsfont}

\allowdisplaybreaks

\usepackage{floatrow}
\newfloatcommand{capbtabbox}{table}[][\FBwidth]

\begin{document}

\title{Power corrections to quasi-distribution amplitudes of a heavy meson}
 \author{Chao Han}
\affiliation{INPAC, Key Laboratory for Particle Astrophysics and Cosmology (MOE),  Shanghai Key Laboratory for Particle Physics and Cosmology, School of Physics and Astronomy, Shanghai Jiao Tong University, Shanghai 200240, China}
\author{Wei Wang} \email{Corresponding author: wei.wang@sjtu.edu.cn}
\affiliation{INPAC, Key Laboratory for Particle Astrophysics and Cosmology (MOE),  Shanghai Key Laboratory for Particle Physics and Cosmology, School of Physics and Astronomy, Shanghai Jiao Tong University, Shanghai 200240, China}
\affiliation{Southern Center for Nuclear-Science Theory (SCNT), Institute of Modern Physics, Chinese Academy of Sciences, Huizhou 516000, Guangdong Province, China}
\author{Jia-Lu Zhang}\email{Corresponding author: elpsycongr00@sjtu.edu.cn }
\affiliation{INPAC, Key Laboratory for Particle Astrophysics and Cosmology (MOE),  Shanghai Key Laboratory for Particle Physics and Cosmology, School of Physics and Astronomy, Shanghai Jiao Tong University, Shanghai 200240, China}
\author{Jian-Hui Zhang}\email{Corresponding author: zhangjianhui@cuhk.edu.cn}
\affiliation{School of Science and Engineering, The Chinese University of Hong Kong, Shenzhen 518172, China}

\begin{abstract}
It has been recently demonstrated that lightcone distribution amplitudes (LCDAs) for a heavy meson, defined in the heavy quark limit, can be extracted by simulating quasi-distribution amplitudes on the lattice with a large meson momentum $P^z$. This extraction involves a two-step procedure. In the first-step, the quasi-distribution amplitudes are matched onto the QCD LCDAs in the large $P^z$ limit. In the second step, the QCD LCDAs are matched onto the desired LCDAs defined in HQET. In this work, we present the  $m_H^{2n}/(P^z)^{2n}$ and $\Lambda_\text{QCD}^2/(xP^z)^2$  and $\Lambda_\text{QCD}^2/((1-x)P^z)^2$corrections in the first step with $m_H$ being the heavy meson mass and $x$ being the momentum fraction of the light quark. To account for $m_H^{2n}/(P^z)^{2n}$ corrections, we employ two methods:  the moment relation that can provide mass corrections to all orders,  and  a leading-twist projector that gives the  leading-order mass corrections.
For the $\Lambda_\text{QCD}^2/(P^z)^2$ corrections, we employ a renormalon model that addresses the renormalon ambiguity stemming from the divergence of the perturbative series. We adopt two parametrizations for QCD LCDAs to conduct a numerical analysis. 
The results suggest that these  corrections  are typically smaller than 20\% in the region $0.2<x<0.8$, and should be included in a precision analysis. 
These findings will serve as a valuable guide for future studies on heavy meson LCDAs, particularly in the context of Lattice QCD, where accounting for such corrections is crucial for improving the accuracy and reliability of the results.
\end{abstract}

\maketitle



\section{Introduction}

Heavy meson lightcone distribution amplitudes (LCDAs) play a crucial role in the realm of heavy flavor physics~\cite{Grozin:1996pq}. In hard exclusive decays of $B$ mesons that are dominated by light-like distances, heavy meson LCDAs emerge in the factorization of decay amplitudes as a convolution with perturbatively-calculable  hard kernels~\cite{Beneke:1999br,Beneke:2000ry,Keum:2000wi,Lu:2000em}. 
The knowledge of LCDAs is essential for making   predictions on decay widths,   and is thus important in the determination of Cabibbo-Kobayashi-Maskawa (CKM) matrix elements as well as in the search for new physics beyond the Standard Model.


Despite the significant progress made in understanding heavy meson LCDAs in recent decades~\cite{Kawamura:2001jm,Lange:2003ff,Braun:2003wx,Lee:2005gza,Charng:2005fj,Kawamura:2008vq,Bell:2013tfa,Feldmann:2014ika,Braun:2019wyx,Galda:2020epp}, accurately modeling them remains a challenge. In the heavy quark effective theory (HQET), heavy meson LCDAs are defined as hadron-to-vacuum matrix elements of nonlocal lightcone operators involving heavy quark fields. Calculating them using first-principle approaches like lattice QCD (LQCD) is extremely difficult, because it seems unlikely to simultaneously realize lightcone separations and heavy quark fields. Furthermore, the existence of cusp divergences also prevents the calculation of non-negative moments of heavy meson LCDAs~\cite{Braun:2003wx}. In QCD sum rules that are widely used for studying heavy meson LCDAs, the available results are limited to the lowest inverse  moments~\cite{Braun:2003wx,Khodjamirian:2005ea,Bell:2013tfa,Meissner:2013hya,Gao:2019lta,Rahimi:2020zzo,Khodjamirian:2023wol}, while phenomenological estimates depend heavily on specific model assumptions~\cite{Bell:2008er,Hwang:2010hw,Yang:2011ie,Wu:2013lga,Sun:2016avp,Binosi:2018rht,Sun:2019xyw,Lan:2019img,Serna:2020txe,Shi:2021nvg,Feldmann:2022uok,Serna:2022yfp,Almeida-Zamora:2023bqb,Arifi:2024mff}.
The lack of precise understanding of heavy meson LCDAs has profound implications. For instance, in predicting form factors for processes like $B\to V$ transitions, the uncertainties stemming from heavy meson LCDAs are dominant~\cite{Gao:2019lta}. These uncertainties introduce unavoidable errors that can affect the determination of CKM matrix elements and undermine the reliability of searches for new physics beyond the Standard Model.

To resolve these difficulties, many interesting explorations have been triggered in the past few years~\cite{Kawamura:2018gqz,Wang:2019msf,Zhao:2019elu,Zhao:2020bsx,Xu:2022krn,Xu:2022guw,Deng:2023csv,Hu:2023bba,Han:2023xbl,Han:2023hgy,Hu:2024ebp,Han:2024ucv}, but no realistic calculation has been achieved  until a two-step matching method was proposed in a recent study~\cite{Han:2024min}. It involves the simulation of an equal-time spatial correlator, called the quasi-distribution amplitudes (quasi-DAs),  of a heavy meson with a large momentum $P^z$, which can be matched onto the QCD LCDAs in the large $P^z$ limit within the large-momentum effective theory (LaMET) framework~\cite{Ji:2013dva,Ji:2014gla,Ji:2020ect}.
The QCD LCDAs are subsequently matched onto the HQET LCDAs in the $m_H\to \infty$ limit~\cite{Ishaq:2019dst,Zhao:2019elu,Beneke:2023nmj,Han:2024min,Han:2024yun,Deng:2024dkd}. This sequential approach lays the foundation for reliable first-principle  predictions 
for heavy meson LCDAs. 
In  this approach, 
a key aspect to ensure the accuracy of the results in the study of heavy meson LCDAs is the inclusion of power corrections. These corrections may have significant impact on the analysis and thus require careful consideration.

In this paper, we investigate the structure and impact of power corrections appearing in the determination of heavy meson LCDAs from their quasi-DA counterparts defined in LaMET. 
To address \( m_H^{2n}/(P^z)^{2n} \) corrections, we employ two distinct methods. The first utilizes the moments relation derived from the operator product expansion (OPE) and provides a formula that incorporates hadron mass corrections of the form \( m_H^{2n}/(P^z)^{2n} \) to all orders~\cite{Zhang:2017bzy,Chen:2016utp}. The second follows from the study of target mass corrections in deep-inelastic scattering (DIS), which enables the calculation of leading-order mass corrections~\cite{Zhang:2017bzy}.
To estimate the $\Lambda_\text{QCD}^2/(P^z)^2$ corrections, we adopt the renormalon model~\cite{Beneke:1994sw,Braun:2018brg,Braun:2024snf} based on the analysis of renormalon ambiguities stemming from the divergence of the perturbative series expansion. For completeness, we also include perturbative $\alpha_s$ corrections in our analysis.  
To test the numerical impact of these corrections, we choose two parametrizations for the LCDA.
The results indicate that for heavy mesons, mass corrections significantly influence the results of the perturbative matching.
These results, together with  power corrections in the matching of QCD LCDAs to HQET LCDAs,  are instrumental in guiding the configuration of lattice simulations for studying heavy meson quasi-DAs in future research endeavors.

The rest of this paper is arranged as follows.
Section 2 provides a brief review of the theory framework. 
Section 3 is dedicated to finite-$P^z$ corrections in the LaMET procedure, including 
$m_H^{2n}/(P^z)^{2n}$ and $\Lambda_\text{QCD}^2/(P^z)^2$  corrections. We use two approaches to calculate $m_H^{2n}/(P^z)^{2n}$ corrections, and then estimate $\Lambda_\text{QCD}^2/(P^z)^2$ corrections using the renormalon model. Subsequently, we outline the procedure for systematically incorporating the power corrections as well as the perturbative corrections.
Section 4 presents a numerical analysis of these corrections. 
A summary is then given in Section 5.
Results for the hard kernel  utilized in the paper are collected in Appendix A.

\section{Theory framework}
\label{sec-th}

The leading-twist LCDAs for a pseudo-scalar heavy meson are defined as
\begin{align}
\label{DA}
 &\phi(x,\mu) = \frac{1}{if_H}\int_{-\infty}^{\infty}\frac{dz}{2\pi}e^{i(x-\frac{1}{2}) z P^+} \notag\\
 &\times \langle 0|\bar{q}(\frac{z n_+}{2})n\!\!\!\slash_+\gamma_5W_c(\frac{z n_+}{2},-\frac{z n_+}{2})Q(-\frac{z n_+}{2})|H(P)\rangle,
\end{align}   
where $n_+^\mu= (1,0,0,-1)/\sqrt{2}$ is a unit lightcone vector. $q(x)$ stands for a light quark and $Q(x)$ stands for a heavy quark. The momentum of the meson is directed along the $n_-^\mu=(1,0,0,-1)/\sqrt{2}$, expressed as $P^\mu=P^{+}  n_-^\mu = (P^z,0,0,P^z)$.
The Wilson line $W_c(x,y)$ is introduced to maintain gauge invariance  whose explicit form is given as
\begin{equation}
W_c(x, y)=\mathcal{P} \exp \left[i g \int_0^1 \mathrm{~d} t(x-y)_\mu A^\mu(t x+(1-t) y)\right].
\end{equation}

The quasi-DAs are defined as
 \begin{align}
\label{quasiDA}
 &\tilde\phi(x,P^z) = \frac{1}{if_H}\int_{-\infty}^{\infty}\frac{dz}{2\pi}e^{-i(x-\frac{1}{2}) z P^z}\tilde{\psi}(z,P^z)\notag\\
  &= \frac{1}{if_H}\int_{-\infty}^{\infty}\frac{dz}{2\pi}e^{-i(x-\frac{1}{2}) z P^z}\notag\\
 &\times \langle 0|\bar{q}(\frac{z n_z}{2})n\!\!\!\slash_z\gamma_5W_c(\frac{z n_z}{2},\frac{-z n_z}{2})Q(\frac{z n_z}{2})|H(P^z)\rangle,
\end{align}   
where $n_z^\mu=(0,0,0,1)$.

After appropriate renormalization, the quasi-DAs can be factorized into a hard kernel and LCDAs through the collinear factorization under LaMET~\cite{Ji:2013dva,Ji:2014gla,Ji:2020ect}: 
\begin{eqnarray} 
\label{matching}
 \Tilde{\phi}_R(x,P^z) &=& \int_0^1 dy C(x,y,\frac{\mu}{P^z})\phi_R(y,\mu) \nonumber\\
 && +\mathcal{O}\left(\frac{m_H^2}{(P^z)^2},\frac{\Lambda_{\mathrm{QCD}}^2}{(P^z)^2}\right),  
\end{eqnarray}
where the subscript $R$ denotes renormalized quantities, $C(x,y,\frac{\mu}{P^z})$ is the perturbative matching kernel, $\mu$ is the factorization scale, and $m_H$ is the mass of the heavy meson.

While LCDAs are conveniently renormalized in the $\overline{\mathrm{MS}}$ scheme, the Lattice QCD community favors the hybrid renormalization scheme for quasi-DAs~\cite{Ji:2020brr,Zhang:2023bxs,Baker:2024zcd}, where one has
\begin{align}
\displaystyle
\tilde{\psi}&_R(z,P^z)=\left\{
\begin{matrix}\displaystyle
\frac{\tilde\psi(z,P^z)}{\tilde\psi(z,0)}\hfill  &  |z|\leq z_s \\
\displaystyle
\frac{\tilde\psi(z,P^z)}{\tilde\psi(z_s,0)}e^{-\delta m(a) (|z|-z_s)} & |z|>z_s
\end{matrix}
\right. .
\label{eq:hybrid_renorm}
\end{align}
$z_s$ is a length scale that separates the short- and long-distance correlations, and $\delta m(a)\sim a^{-1}$ is a mass counterterm that removes the linear divergence associated with the Wilson line self energy.
{By dividing the bare quasi-DA by a chosen zero-momentum matrix element $\tilde\psi(z,0)$, the UV divergences can be removed properly without introducing additional long-distance effects~\cite{Ji:2020brr}, and can be implemented on the lattice~\cite{Zhang:2023bxs,Baker:2024zcd}. }

As can be seen from Eq.~(\ref{matching}), the difference between the quasi-DAs and LCDAs at leading-power accuracy is captured by the matching kernel $C$. Its explicit expression in the hybrid scheme is given in Appendix A.
{It is important to note that the quark mass correction is not included in the matching kernel. To diminish such effects that may play a crucial role in determining the LCDAs, one should increase the hadron momentum in lattice simulations. On phenomenological side, the inclusion of quark mass would also help to improve the accuracy. This   will be investigated in our future work.} In addition to quark mass correction, there are  the power corrections of ${\cal O}(\frac{m_H^2}{(P^z)^2},\frac{\Lambda_{\mathrm{QCD}}^2}{(P^z)^2})$ in Eq.~(\ref{matching}), which for light mesons might not  have significant impact except in the endpoint region. However, for heavy mesons, due to their large mass and the finiteness of $P^z$ in practical lattice simulations, it is essential to estimate the impact of these power corrections, in order to obtain precise results for heavy meson LCDAs. 

\section{Finite-$P^z$ Power corrections}
\label{sec-power}


\subsection{$m_H^{2n}/(P^z)^{2n}$ correction}

The origin of hadron mass corrections can be observed using the OPE. One can write down the inverse Fourier transform of Eq.~(\ref{DA}): 
\begin{align}
\label{inversfourier}  
    \langle 0|\bar{q}(\frac{z n_+}{2})n\!\!/_+\gamma_5&W_c(\frac{z n_+}{2},-\frac{z n_+}{2})Q(-\frac{z n_+}{2})|H(P)\rangle \notag=\\
if_HP^+&\int_{0}^{1}dx e^{-i(x-\frac{1}{2})z P^+}\phi(x,\mu).
\end{align}
Expanding both sides of Eq.~(\ref{inversfourier}) gives
\begin{align}
        \label{ope1}
    &if_H(P^+)^{n+1}\int_{0}^{1}dx (1-2x)^n\phi(x,\mu)\notag\\
    &=\langle 0|\bar{q}(0)n\!\!/_+(i{\mathop{D}\limits^\leftrightarrow}_+)^n\gamma_5Q(0)|H(P)\rangle\notag\\
    &=i^n n_{+}^{\mu_1}...n_{+}^{\mu_n}\langle 0|\bar{q}(0)\gamma_{\mu_1}{\mathop{D}\limits^\leftrightarrow}_{\mu_1}...{\mathop{D}\limits^{\leftrightarrow}}_{\mu_n}\gamma_5Q(0)|H(P)\rangle\notag\\
        &=i^n \langle\langle O^n\rangle\rangle n_{+}^{\mu_1}...n_{+}^{\mu_n}P_{\mu_1}...P_{\mu_n}\notag\\
        &=i^n \langle\langle O^n\rangle\rangle n_{+}^{(\mu_1}...n_{+}^{\mu_n)}P_{(\mu_1}...P_{\mu_n)},
\end{align}
where ${\mathop{D}\limits^{\rightarrow}}_{\mu_n}={\mathop{\partial}\limits^{\rightarrow}}_{\mu_n}-igA_\mu$, ${\mathop{D}\limits^{\leftarrow}}_{\mu_n}={\mathop{\partial}\limits^{\leftarrow}}_{\mu_n}+igA_\mu$, ${\mathop{D}\limits^{\leftrightarrow}}_{\mu_n}={\mathop{D}\limits^{\leftarrow}}_{\mu_n}-{\mathop{D}\limits^{\rightarrow}}_{\mu_n}$, $(\mu_1...\mu_n)$ stands for the symmetric and traceless combination of the enclosed indices, and $\langle\langle O^n\rangle\rangle$ is the reduced matrix element of leading-twist operators. 
When considering the LCDAs and quasi-DAs, the reduced matrix elements are corresponding to the moments.
The last equality holds because, since $n_+^2=0$, only the symmetric and traceless part of $P^{\mu_1}...P^{\mu_n}$ contributes.
The moment is defined by
\begin{eqnarray}
\langle (1-2x)^{n-1}\rangle &=&\displaystyle\int_{-\infty}^{\infty}dx (1-2x)^{n-1}\phi(x).
\end{eqnarray}
However, in the case of quasi-DAs, the trace part also contributes, introducing mass corrections compared to LCDAs. 
This contribution results in a modification of the form to \(n_{z}^{\mu_1} \ldots n_{z}^{\mu_n} P_{\mu_1} \ldots P_{\mu_n}\).
The discrepancies between the moments of the LCDA and those of the quasi-DA are encapsulated by the relation
\begin{align}
    \frac{\langle x^{n-1}\rangle_{\tilde\phi}}{\langle x^{n-1}\rangle_\phi}=\frac{n_{(\mu_1}...n_{\mu_n)}P^{\mu_1}...P^{\mu_n}}{n_{\mu_1}...n_{\mu_n}P^{\mu_1}...P^{\mu_n}}=\sum_{i=0}^{i_{\text{max}}}C_{n-i}^ic^i,
\end{align}
where $\tilde \phi$ stands for the quasi-DA, $C_{n-i}^i$ is the binomial function and $c=-n^2 m_H^2/(4 (n\cdot P)^2)=m_H^2/(4 P^z)^2$. 
Subsequently, by following the procedure outlined in~\cite{Zhang:2017bzy, Chen:2016utp,Ma:2017pxb}, the mass corrections applicable to LCDAs can be written out directly
\begin{equation}
    \begin{aligned}
    \label{mc1}
        {\phi_{I}}&(x,P^z,\mu)=\frac{1}{\sqrt{1+4c}} \\
        &\times \bigg[\frac{f_+}{2}\phi(\frac{1}{2}-\frac{1-2x}{f_+},\mu)-\frac{f_-}{2}\phi(\frac{1}{2}+\frac{1-2x}{f_-},\mu)\bigg],
    \end{aligned}
\end{equation}
where $f_\pm=\sqrt{1+4c}\pm 1$ and $\phi_I(x)$ stands for the mass-corrected LCDAs.

An alternative method to calculate the hadron mass correction is given in Ref.~\cite{Braun:2018brg}. Note that through lightcone OPE~\cite{Anikin:1978tj,Anikin:1979kq,Balitsky:1987bk,Muller:1994ses,Balitsky:1990ck,Geyer:1999uq}, the LCDAs can be defined via
 \begin{equation}
     \begin{aligned}
     \label{mc2}
\phi_{II}&(z,P^z,\mu)=\langle 0|\Pi_{l.t}^{\mu}[\bar{q}(\frac{zn_z}{2}) {} n\!\!\!\slash_z\gamma_5Q(-\frac{zn_z}{2})]|H(P^z)\rangle\\
=&if_H\int_0^1dx \Pi_{l.t}[(P\cdot n_z)e^{-i(x-1/2)zP\cdot n_z}]\phi(x,\mu)\\
=&if_H\int_0^1dx (P^z-\frac{i}{8}(2x-1)m_H^2z)
\\&
\times e^{i(x-1/2)zP^z} \phi(x,\mu)+\mathcal{O}(\frac{m_H^4}{(P^z)^4})\\
=&if_HP^z\int_0^1dx (1+\frac{1}{8}(2x-1)\frac{m_H^2}{(P^z)^2}\frac{d}{dx})\\
&\times e^{i(x-1/2)zP^z}\phi(x,\mu)+\mathcal{O}(\frac{m_H^4}{(P^z)^4})\\
=&if_HP^z\int_0^1dx e^{i(x-1/2)zP^z} \bigg (\phi(x,\mu)-\frac{1}{8}\frac{m_H^2}{(P^z)^2}\\
&\times \frac{d}{dx}((2x-1)\phi(x,\mu))\bigg )+\mathcal{O}(\frac{m_H^4}{(P^z)^4}),
     \end{aligned}
 \end{equation}
 where $\Pi_{l.t}^{\mu}$ stands for the leading-twist projection. 
 It should be mentioned that this method only provides the leading-power mass correction. We have verified that it agrees with Eq.~(\ref{mc1}) when the latter is expanded to the same order.

 
Technically speaking, as an off-forward matrix element, LCDA also receives mass corrections from higher-twist matrix elements that can be reduced to the total derivatives of twist-2 operators~\cite{Ball:1998kk,Braun:2011zr,Braun:2011dg}, which is not included in the above discussion.
Explicitly, these operators have the form of
\begin{equation}
\mathcal{O}_1=\partial^2 \mathcal{O}_{\mu_1 \ldots \mu_n}, \quad \mathcal{O}_2=\partial^{\mu_1} \mathcal{O}_{\mu_{1 \ldots} \ldots \mu_n},
\end{equation}
where the $\mathcal{O}_{\mu_1 \ldots \mu_n}$ is a symmetric and traceless twist-2 operator.
Given that these are merely higher-twist terms, we choose to neglect them in the mass correction sector but incorporate these effects when we discuss $\frac{\Lambda_{\mathrm{QCD}}^2}{(P^z)^2}$ corrections. 

\subsection{$\Lambda_\text{QCD}^2/(P^z)^2$ Correction}

In Eq.~(\ref{matching}), higher-twist operators also contribute to the quasi-DAs, and belong to power corrections of the form $\frac{\Lambda_{\mathrm{QCD}}^{2n}}{(P^z)^{2n}}$, where $n$ is an integer.  There are multiple ways to consider these higher-twist contributions. The higher-twist operator matrix elements have been given for the parton distribution function (PDF) and pion LCDAs~\cite{Zhang:2017bzy,Chen:2016utp}, but they are difficult to calculate on the lattice. On the other hand, the renormalon model has been used to estimate the higher-twist contributions to the PDF and the generalized parton distribution (GPD)~\cite{Braun:2018brg,Braun:2024snf}. In this paper, we employ the renormalon model to examine the higher-twist contribution to the heavy meson LCDAs.

The spirit of the renormalon model can be understood as follows. 
When employing cutoff regularization in the factorization formula (Eq.~(\ref{matching})), the matching kernel takes the form
\begin{equation}
C(x,y,\frac{\mu_F}{P^z})=\delta(x-y)+\sum_{i=1}^{\infty} c_i\alpha_s^i-\frac{\mu_F^2}{(P^z)^2}D_{\mathcal{Q}}(x),
\end{equation}
where $\mu_F$ is the cutoff scale, and $c_i$ depend logarithmically on  $\mu_F$. The  $D_{\mathcal{Q}}(x)$ term represents the leading power correction to the matching kernel. Since the left-hand side of Eq.~(\ref{matching}) is independent of the factorization scale, the power correction term must cancel with those arising from higher-twist operators, and thus indicating the magnitude of higher-twist contributions.

However, in the context of dimensional regularization, a more commonly used regularization in practical calculations, the power corrections do not manifest themselves. Instead, the perturbative series $\sum_{i=1}^{\infty} c_i\alpha_s^i$ diverge asymptotically with a zero convergence radius. 
To address this issue, Borel summation techniques can be employed to transform this asymptotic series into the Borel plane. 
For the matching kernel, the Borel transform takes the following form 
\begin{equation}
B[T](w)=\sum_{k=1}^{\infty} \frac{c_k}{k!}\left(\frac{w}{\beta_0}\right)^k.
\end{equation}
The Borel integral
\begin{equation}
\label{borel}
T(\alpha_s)=\frac{1}{\beta_0} \int_0^{\infty} d w e^{-w /\left(\beta_0 a_s\right)} B[T](w)
\end{equation}
then yields the asymptotic series $\sum_{i=1}^{\infty} c_i\alpha_s^i$ given above. However,
this integral is not well-defined as it has singularities on the integration path. One can choose a contour to circumvent the singularities. But the choice is arbitrary, and introduces the so-called renormalon ambiguity. 
The renormalon model is based on the expectation that the ambiguity is of the same order as the contribution from higher-twist operators, which allows the ambiguity to be used in parameterizing these higher-twist contributions.

\begin{@twocolumnfalse}
    \begin{center}
    \begin{figure*}
\label{pic_model}
\centering
\begin{minipage}[t]{0.3\textwidth}
\centering
\includegraphics[width=5cm]{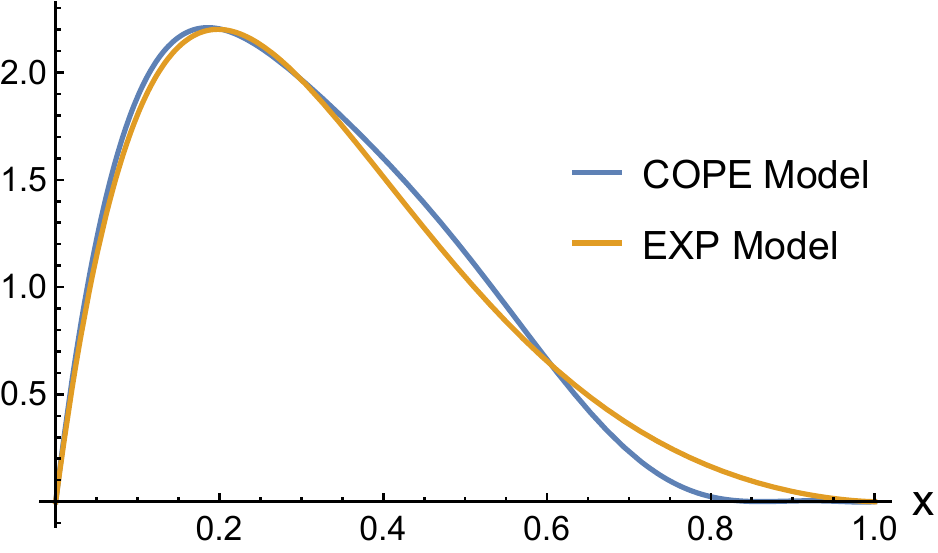}
\centerline{(a)}
\end{minipage}
\begin{minipage}[t]{0.3\textwidth}
\centering
\includegraphics[width=5cm]{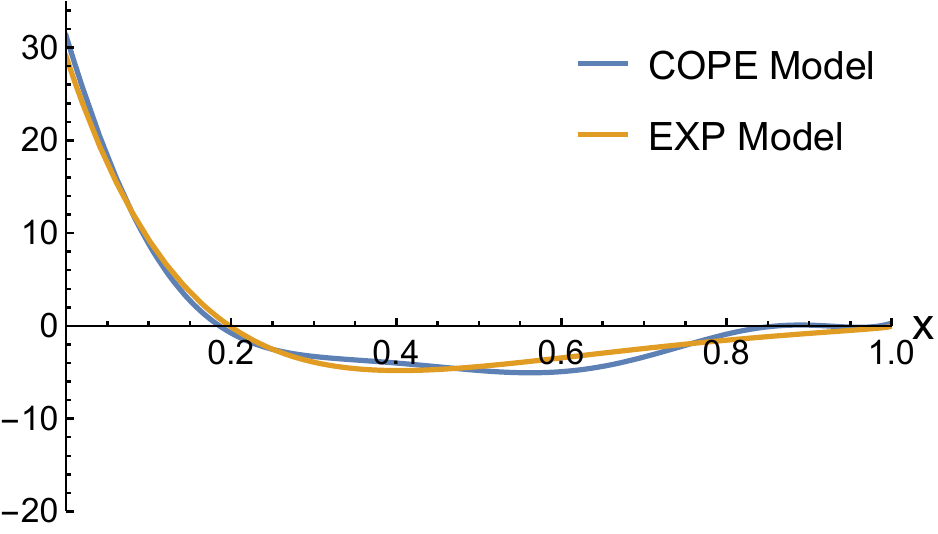}
\centerline{(b)}
\end{minipage}
\begin{minipage}[t]{0.3\textwidth}
\centering
\includegraphics[width=5cm]{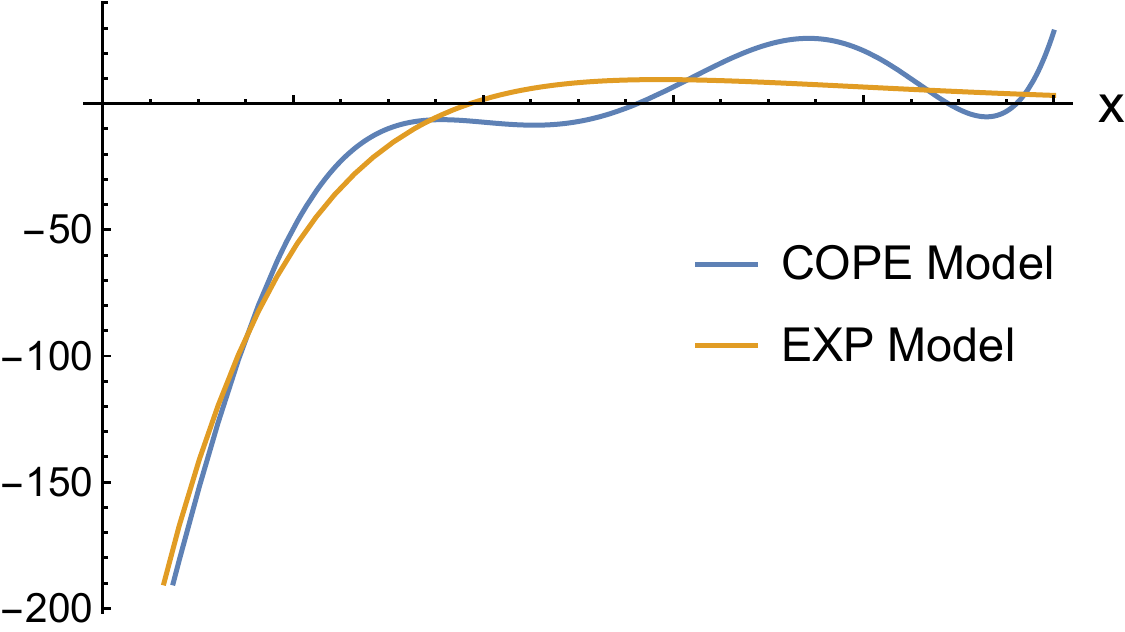}
\centerline{(c)}
\end{minipage}
\caption{(a) The two LCDA models used in this study. (b) The first derivative of these two models. (c) The second derivative of these two models.}
\label{model}
\end{figure*}
    \end{center}
\end{@twocolumnfalse}

In the Borel plane, the first singularity occurs at $w=1/2$, which is a UV renormalon associated with the self-energy of the gauge link in the quasi-DAs. Similar to the heavy quark renormalon, this UV renormalon has been studied in~\cite{Beneke:1994sw}. 
UV renormalons are insensitive to the IR physics. 
They are cancelled out through renormalization. 

The twist-4 operators correspond to the singularity at $w=1$, which represents an IR renormalon effect. Based on this, we use the renormalon model to estimate the higher-twist contribution. The ambiguity  is calculated in~\cite{Braun:2024snf} and equals $\mathcal{N}\frac{m_H^2}{(P^z)^2}\delta_R\Phi(x,\mu)$, where
\begin{equation}
\begin{aligned}
\label{ambiguity}
& 2 \delta_R \Phi(x,\mu)
= 
\left[\frac{\Phi'(x,\mu)}{x-1}  \right.
\\&
-  \frac{1}{(x-1)^2} \int_{-1}^{x}\!dy\, \left[\frac{x-1}{y-1}+\ln\left(1-\frac{x-1}{y-1}\right)\right] \, \Phi'(y,\mu)\bigg]  
\\&\quad
+ \left[ \frac{\Phi'(x,\mu)}{x+1} \right.
\\&
+ \frac{1}{(x+1)^2} \int_{x}^1\!dy\,\left[\frac{x+1}{y+1}+\ln\left(1-\frac{x+1}{y+1}\right)\right] \, \Phi'(y,\mu)\bigg] 
\\&+\frac{1}{4}\Phi''(x,\mu),
\end{aligned}
\end{equation}
and $\mathcal{N}=\displaystyle\frac{C_F e^{5/3}}{\beta_0}\sim 0.76$.
{Due to a different convention of Fourier transformation, our DA $\phi(x,\mu)$ is different from but can related to $\Phi(x,\mu)$  through $\Phi(1-2x)=\phi(x)$.} The second derivative term \(\Phi''(x,\mu)\) arises as a consequence of employing the hybrid renormalization scheme.

\subsection{Implementation of power corrections}
\label{sub-order}

As discussed in Section \ref{sec-th}, the finite value of $P^z$, with a maximal value in the range of $3\,\mathrm{\sim}\,5~\mathrm{GeV}$ in practical lattice calculations, urges one to include the power corrections.
We now need to establish a proper sequence for applying various corrections.

Actually, the power corrections can be included through the factorization in coordinate space: 
\begin{equation}
\begin{aligned}
\label{CLaMET}
    &\langle 0|\bar{q}(\frac{z n_+}{2})n\!\!/_+\gamma_5Q(-\frac{z n_+}{2})|H(P^z)\rangle=\int_{0}^1 d\alpha \mathcal{H}(z,\alpha,\mu_F)\\
    &\times \langle 0 |\Pi_{l.t.}^{\mu_F}[\bar{q}(\frac{z n_+}{2})n\!\!/_+\gamma_5Q(-\frac{z n_+}{2})]|H(P^z)\rangle+\mathcal{O}_{twist-4}+...,
\end{aligned}
\end{equation}
where 
$\mathcal{H}(z,\alpha,\mu_F)$ is the matching kernel in coordinate space, and $\mathcal{O}_{twist-4}+...$ stands for higher-twist contributions.

Combining Eq.~(\ref{mc2}) and Eq.~(\ref{CLaMET}) provides a proper sequence for incorporating power corrections. In this paper, we begin by including mass corrections to the LCDAs, followed by performing one-loop matching to obtain the quasi-DAs. Finally, we incorporate the renormalon model estimate of twist-4 corrections into the quasi-DAs. 

\begin{@twocolumnfalse}
    \begin{center}
    \begin{figure*}[htbp]
\centering
\begin{minipage}[t]{0.4\textwidth}
\centering
\includegraphics[width=7cm]{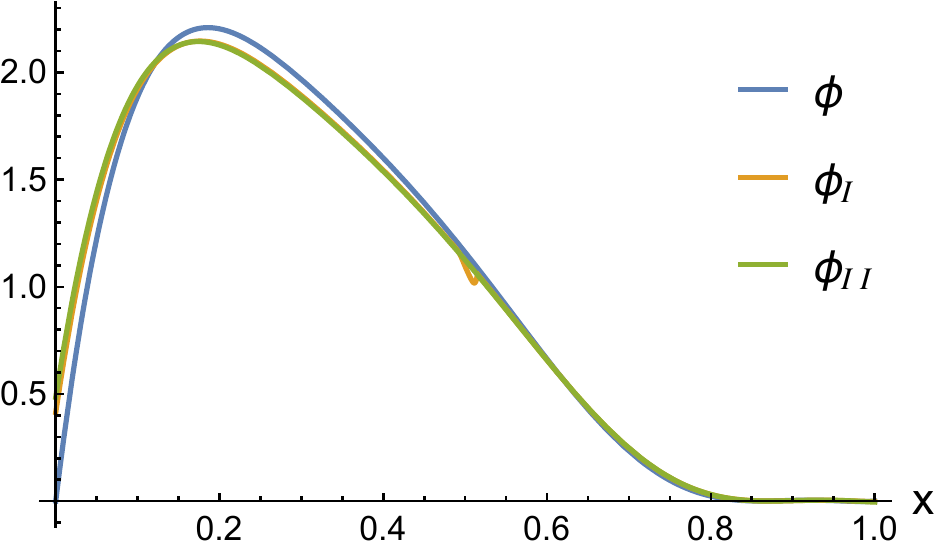}
\centerline{(a)}
\end{minipage}
\begin{minipage}[t]{0.5\textwidth}
\centering
\includegraphics[width=7cm]{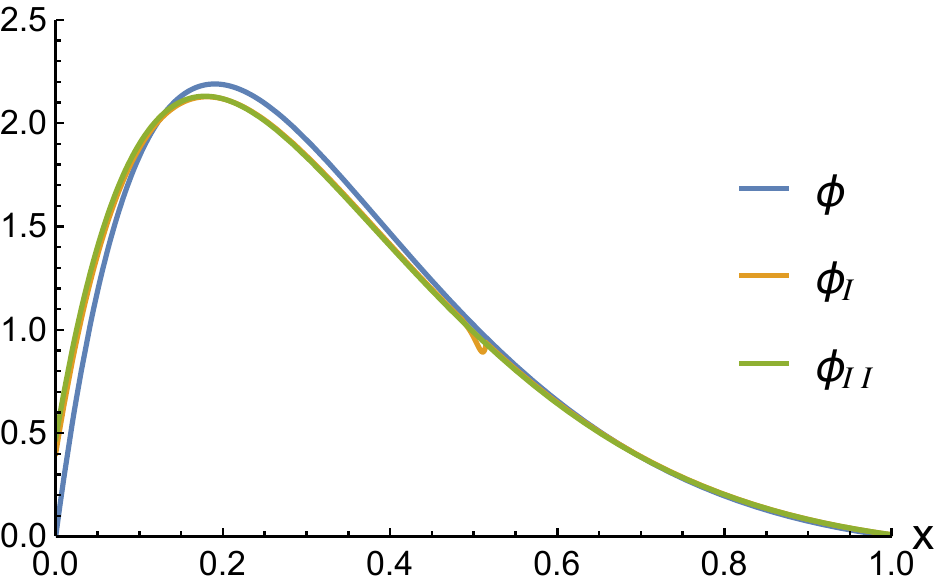}
\centerline{(b)}
\end{minipage}
\begin{minipage}[t]{0.4\textwidth}
\centering
\includegraphics[width=7cm]{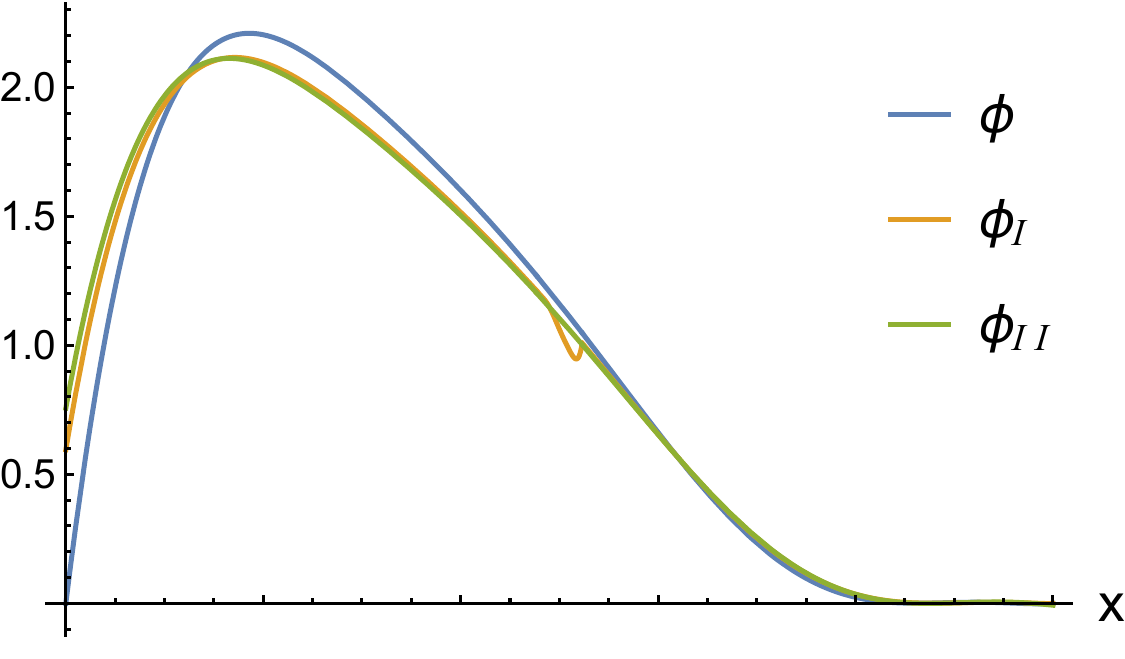}
\centerline{(c)}
\end{minipage}
\begin{minipage}[t]{0.5\textwidth}
\centering
\includegraphics[width=7cm]{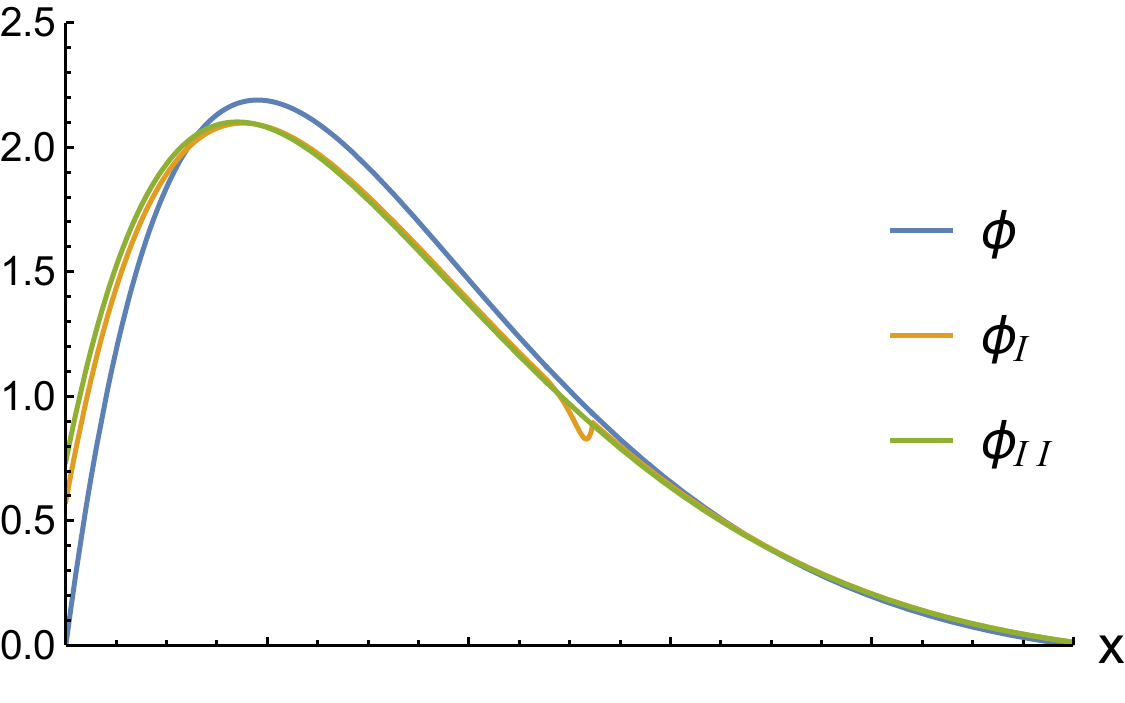}
\centerline{(d)}
\end{minipage}
\begin{minipage}[t]{0.4\textwidth}
\centering
\includegraphics[width=7cm]{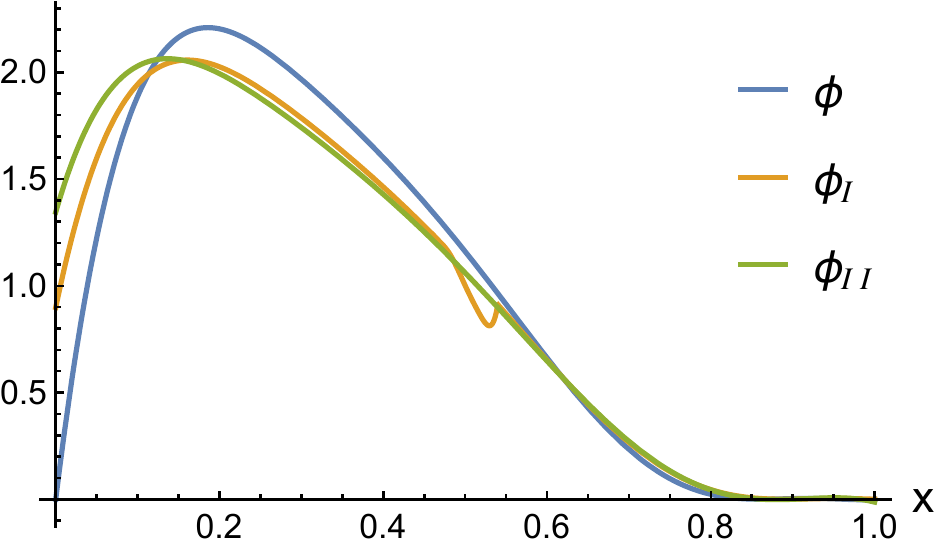}
\centerline{(e)}
\end{minipage}
\begin{minipage}[t]{0.5\textwidth}
\centering
\includegraphics[width=7cm]{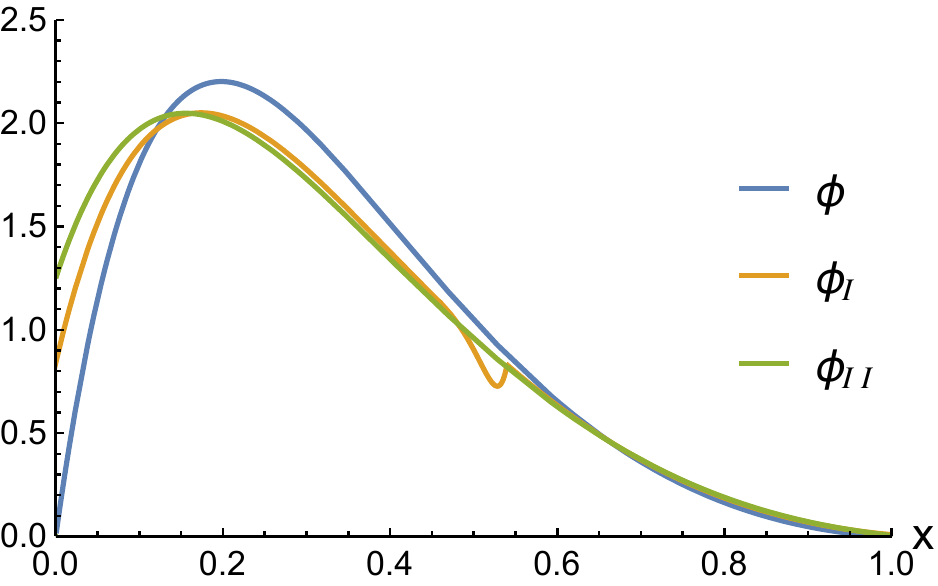}
\centerline{(f)}
\end{minipage}
\caption{Mass corrections to the LCDAs at $\mu=1.6~\mathrm{GeV}$. 
For the subfigures labeled by (a) and (b), we have chosen \(P^z=5~\mathrm{GeV}\); for the subfigures labeled by (c) and (d), \(P^z=4~\mathrm{GeV}\); and for those labeled by (e) and (f), \(P^z=3~\mathrm{GeV}\). 
The subfigures labeled by (a) , (c) and (e), represent the mass correction to the COPE model, while the subfigures labeled by (b), (d) and (f),  represent the mass correction to the EXP model.
$\phi$ is the original DA, and $\phi_I$ is mass-corrected through Eq.~(\ref{mc1}) and $\phi_{II}$ is mass-corrected through Eq.~(\ref{mc2}).}
\label{pic-mass}
\end{figure*}
    \end{center}
\end{@twocolumnfalse}

\begin{@twocolumnfalse}
    \begin{center}
        \begin{figure*}[htbp]
\centering
\begin{minipage}[t]{0.4\textwidth}
\centering
\includegraphics[width=7cm]{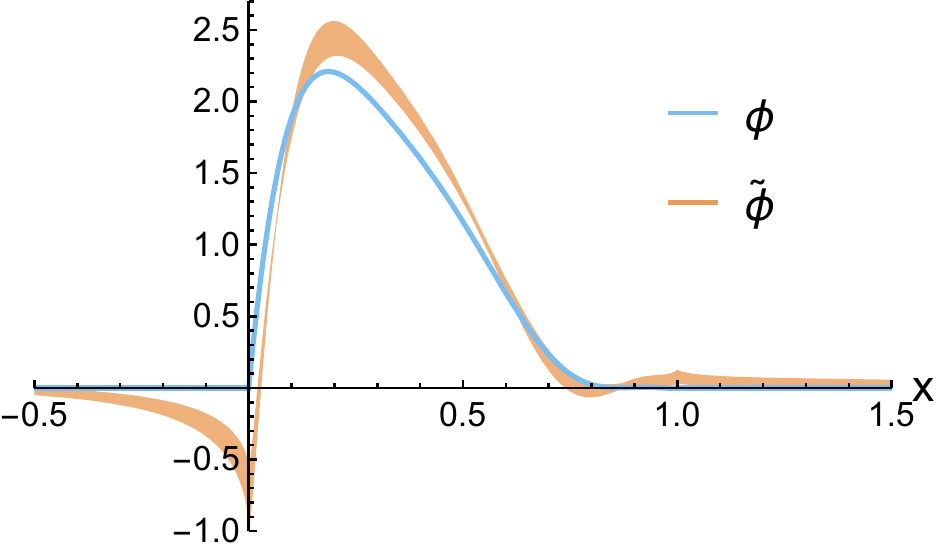}
\centerline{(a)}
\end{minipage}
\begin{minipage}[t]{0.5\textwidth}
\centering
\includegraphics[width=7cm]{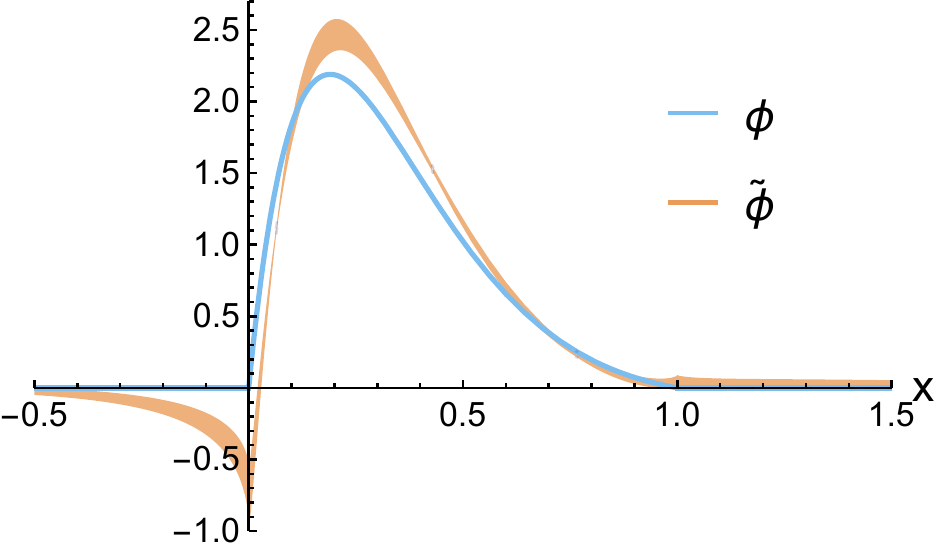}
\centerline{(b)}
\end{minipage}

\caption{These figures show one-loop correction to LCDAs at $P^z=3~\mathrm{GeV}$. (a) and (b) are for the COPE model and the EXP model, respectively. 
In both panels (a) and (b), \( \phi \) represents the LCDA, and \( \tilde{\phi} \) stands for the one-loop-corrected quasi-DA.}
\label{loop-pic}
\end{figure*}
    \end{center}
\end{@twocolumnfalse}

\begin{@twocolumnfalse}
    \begin{center}
        \begin{figure*}[htbp]
\centering
\begin{minipage}[t]{0.4\textwidth}
\centering
\includegraphics[width=7cm]{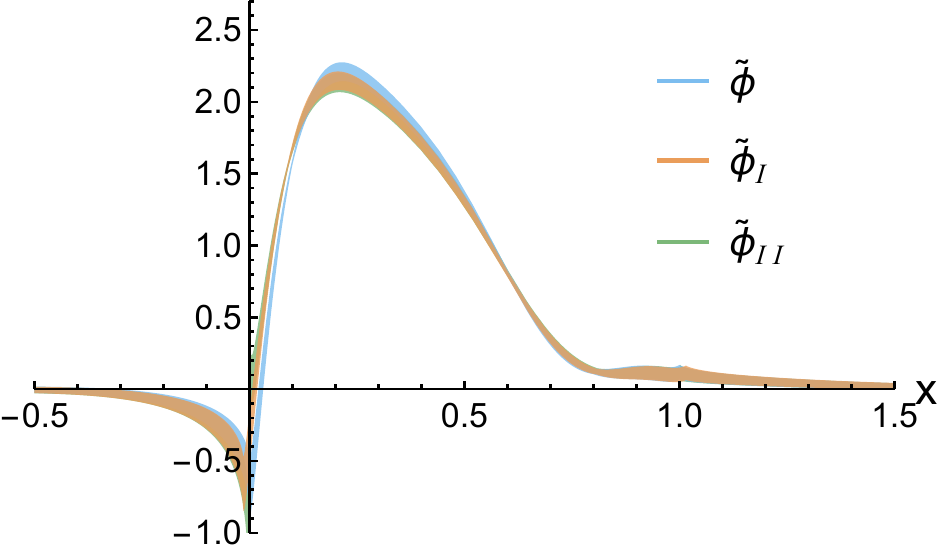}
\centerline{(a)}
\end{minipage}
\begin{minipage}[t]{0.5\textwidth}
\centering
\includegraphics[width=7cm]{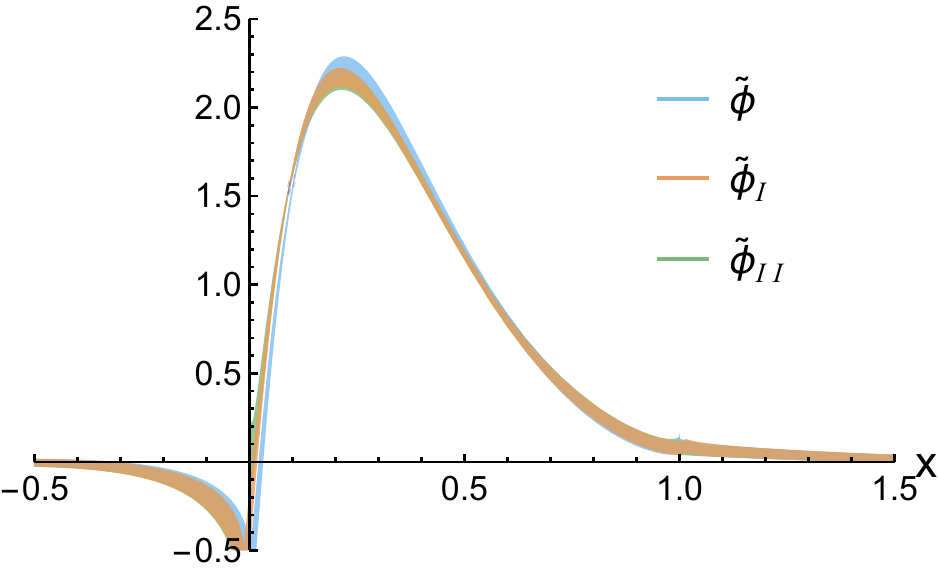}
\centerline{(b)}
\end{minipage}
\begin{minipage}[t]{0.4\textwidth}
\centering
\includegraphics[width=7cm]{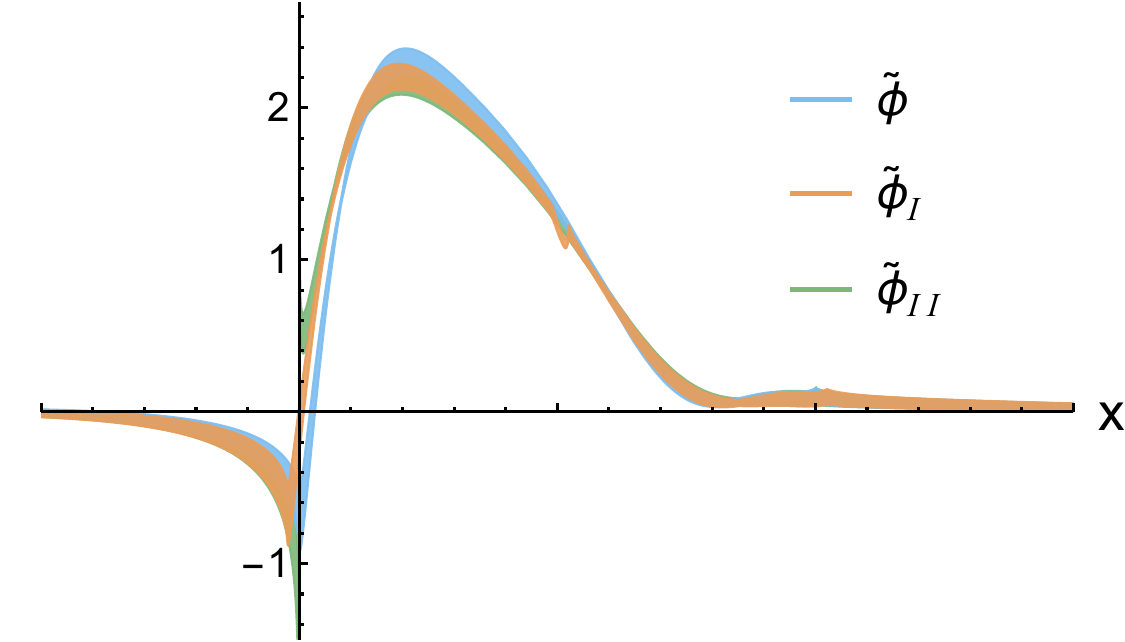}
\centerline{(c)}
\end{minipage}
\begin{minipage}[t]{0.5\textwidth}
\centering
\includegraphics[width=7cm]{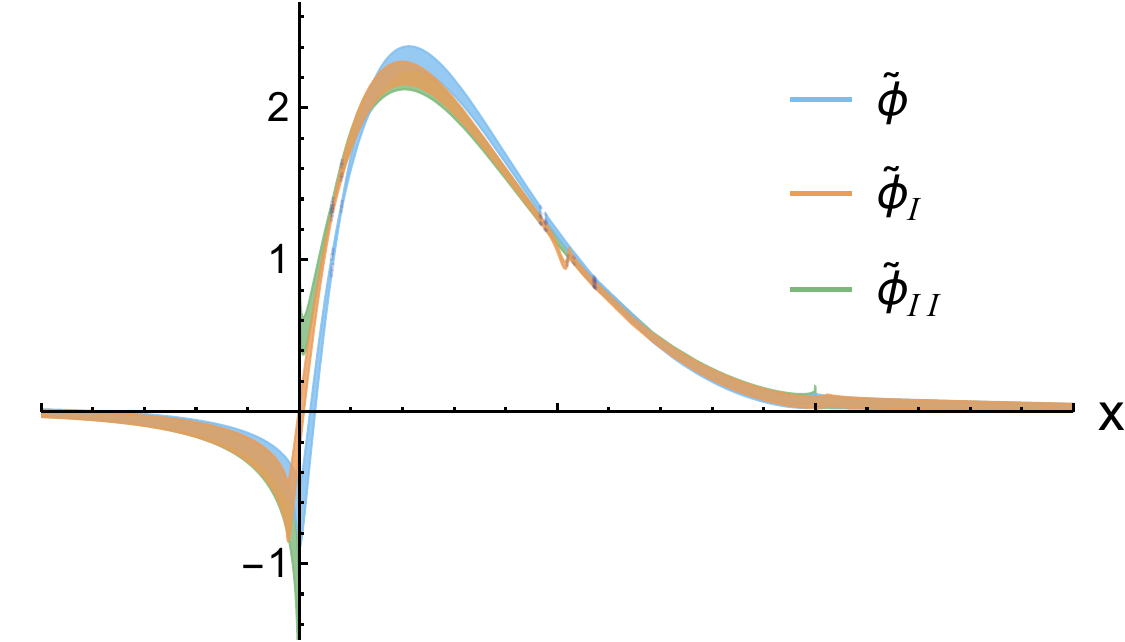}
\centerline{(d)}
\end{minipage}
\begin{minipage}[t]{0.4\textwidth}
\centering
\includegraphics[width=7cm]{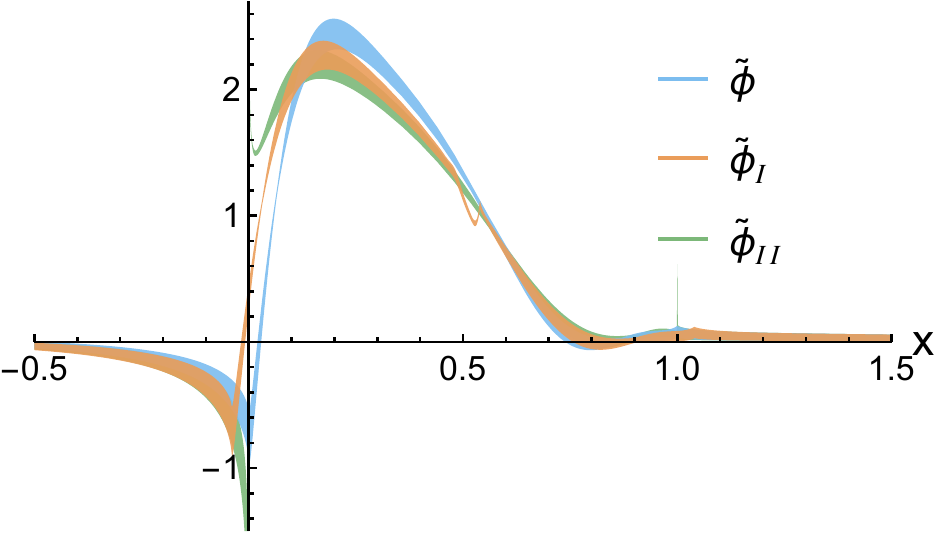}
\centerline{(e)}
\end{minipage}
\begin{minipage}[t]{0.5\textwidth}
\centering
\includegraphics[width=7cm]{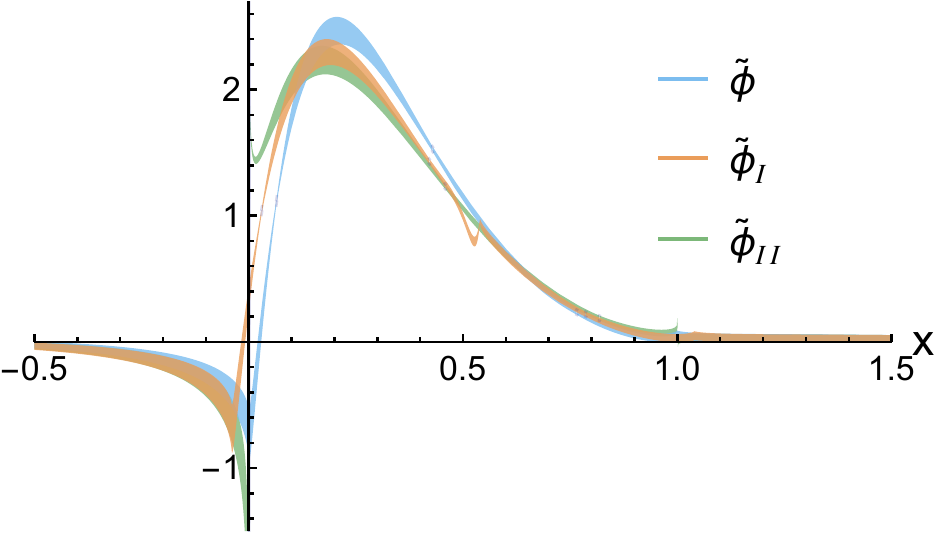}
\centerline{(f)}
\end{minipage}
\caption{These figures show the mass corrections to quasi-DAs. 
The subfigures labeled by (a) and (b), correspond to \(P^z=5~\mathrm{GeV}\), labeled by (c) and (d) to \(P^z=4~\mathrm{GeV}\), and labeled by (e) and (f) to \(P^z=3~\mathrm{GeV}\). 
The subfigures (a), (c) and (e) represent the mass correction to the COPE model, while the subfigures (b), (d) and (f) represent the mass correction to the EXP model.
$\tilde\phi$ is the quasi-DA with only one-loop correction. $\tilde\phi_I$ is mass-corrected through Eq.~(\ref{mc1}). $\tilde\phi_{II}$ is mass-corrected through Eq.~(\ref{mc2}). }
\label{loop-mass-pic}
\end{figure*}
    \end{center}
\end{@twocolumnfalse}

\section{Numerical analysis}
Following the procedure outlined in the previous section, we present the numerical analysis on power corrections by taking the $D$-meson LCDA as an example. We consider two LCDA models and then investigate the impact of power corrections when converting the results to quasi-DAs. Our choice of the relevant parameters is 
\begin{equation}
\begin{aligned}
m_H=1.76&~\mathrm{GeV},\Lambda_{\text{QCD}}=0.5~\mathrm{GeV},z_s=0.1672~\mathrm{fm},\\&\mu=1.6\sim 3~\mathrm{GeV}, P^z=\{3, 4,5\}~\mathrm{GeV}.
\end{aligned}
\end{equation}



\subsection{Model parametrizations}
The first LCDA model used here is given by the conformal moments \cite{Beneke:2023nmj} as
\begin{equation}
\label{}
\phi(x,\mu) = 6x(1-x)\biggl[1+\sum_{n=1}^{\infty}a_n(\mu)C^{(3/2)}_n(2x-1) \biggr]\,,
\end{equation}
where $C_n^{(3/2)}(2x-1)$ are the Gegenbauer polynomials. We truncate the sum to $n=6$ with
\begin{equation}
a_n^D(\mu_c) = \{-0.659, 0.206, -0.057, 0.036, -0.004, -0.007\}\,.
\end{equation}
This model is referred to as the ``COPE model" in this section. 
However, it should be noticed that using the COPE model has certain drawback. 
Just as the Fourier transform, the conformal OPE expansion contains increasingly oscillating terms. This parametrization gives a smooth shape for LCDA itself, but the corresponding  derivatives of LCDA highly oscillate.  
This will introduce additional effects, as can be seen in subsequent discussions. 

Since the power corrections heavily depend on the first and second derivatives of the LCDAs, we also use another model which is relatively smooth     
\begin{equation}
\phi(x,\mu)=a(1-x)x e^{-bx},
\end{equation}
where the parameters $a,b$ are fitted to the  LCDAs in the ``COPE model" with $a=31.451$, $b=-4.022$ at $\mu=1.6~\mathrm{GeV}$, and $a=25.670$, $b=-3.493$ at $\mu=3~\mathrm{GeV}$. 
The second model is referred to as ``EXP model" in this section. 

The shape of two LCDA models and their derivatives are shown in Fig.~\ref{model}.
Both models show that the hadron momentum is predominantly carried by the heavy quark. 
It should be noted that the quasi-DAs are not confined to the range of [0,1], but extend over a broader range. Actually, the smaller the momentum $P^z$ is, the wider the quasi-DAs are.
As the boosted momentum increases, the quasi-DA range gradually converges to the LCDA range.

\subsection{{$m_H^{2n}/(P^z)^{2n}$ Corrections}}

Following the procedure outlined in Section~\ref{sub-order}, we first consider the $m_H^{2n}/(P^z)^{2n}$ corrections, which are introduced through two methods.
The mass-corrected DA through Eq.~(\ref{mc1}) is referred to as $\phi_{\uppercase\expandafter{\romannumeral1}}$ while that through Eq.~(\ref{mc2}) as $\phi_{{\uppercase\expandafter{\romannumeral2}}}$.
In Fig.~\ref{pic-mass}, we show a comparison of the mass-corrected results obtained by applying the two methods to the LCDA. 
{Since the majority of the momentum fraction is carried by the charm quark, our primary focus should be on the region 
$x \to 0$}.\footnote{When considering the target mass correction, the region of interest depends on the type of DA. For a symmetric DA, such as that of pion or charmonium, the effects are expected to be similar in both the \(x \to 0\) and \(x \to 1\) regions. In contrast, for a \(\bar{D}\) meson, the focus shifts toward the \(x \to 1\) region.
}
As expected, the mass correction has a significant impact on the results due to the large meson mass. 
The mass-corrected distribution becomes wider, and the peaks are lower and slightly shifted towards $x=0$.
There is also an enhancement at the $x\to 0$ endpoint region.  
{
This can be understood by observing that there is a derivative of the DA in Eq.~(\ref{mc2}). From the first derivatives of the two models shown in Fig.~\ref{model}, we can anticipate that they will yield significantly larger corrections in the \( x \to 0 \) region.
}

We also observe a significant difference between the two mass correction methods, highlighting the distinctions between the all-order mass correction and the leading-order correction. 
The DA corrected through Eq.~(\ref{mc1}) exhibits a dent at $\displaystyle x=\frac{1}{2}$, whereas the DA corrected through Eq.~(\ref{mc2}) is relatively smooth. 
This small dent originates from the term $\displaystyle\frac{f_-}{2}\phi(\frac{1}{2}+\frac{1-2x}{f_-},\mu)$, whose amplitude and support are both suppressed by the small parameter $f_-\sim 0.17$. 
Consistency between the two mass corrections can be established by performing a Taylor expansion of the all-order mass correction and extracting the leading-order results, which align with the leading-order correction obtained using the projector method. 
Notably,  at leading power in  $m_H/P^z$ the term  $\displaystyle\frac{f_-}{2}\phi(\frac{1}{2}+\frac{1-2x}{f_-},\mu)$ is zero.   

The correction through Eq.~(\ref{mc2}) is larger than that through Eq.~(\ref{mc1}), especially in the $x\to 0$ region, even considering that the correction from the former is only $m_H^{2}/(P^z)^{2}$ while the corrections from latter include all $m_H^{2n}/(P^z)^{2n}$ contributions. 
This is due to the fact that there are also distributions for $\phi_I$ at the negative axis.

Additionally, when $P^z=5~\mathrm{GeV}$, the mass correction is relatively small, and the two mass correction methods are similar apart from the dent at $x=\frac{1}{2}$. 
However, as the momentum \( P^z \) decreases, particularly when \( P^z = 3~\mathrm{GeV} \), the mass correction is quite significant, and the results of the two methods deviate more noticeably at small $x$.

\subsection{${\cal O}(\alpha_s)$ corrections}

We now proceed to incorporate the ${\cal O}(\alpha_s)$ correction into the preceding discussions. 
Firstly, we consider quasi-DAs with only one-loop correction.  
Subsequent to the application of this kernel, the results for the \( P^z=3~\mathrm{GeV} \) case are depicted in Fig.~\ref{loop-pic}. 
{The error bands in Fig.~\ref{loop-pic} and Fig.~\ref{loop-mass-pic} arise from the variation of the factorization scale, ranging from $1.6~\mathrm{GeV}$ to $3~\mathrm{GeV}$.}
One can observe that the one-loop correction narrows the distribution, raises the peak, and shifts it toward the middle, compared to the original tree-level results.

Subsequently, we consider quasi-DAs with both mass correction and  one-loop correction.
Figure~\ref{loop-mass-pic} displays the corresponding both mass-corrected and one-loop-corrected quasi-DAs at  $P^z=3~\mathrm{GeV}$, $P^z=4~\mathrm{GeV}$, and $5~\mathrm{GeV}$, respectively. 
The dent in the results incorporating mass corrections, as governed by Eq.~(\ref{mc1}), is amplified by the one-loop correction. 
Near $x \rightarrow 1$, the distributions are lifted up, whereas in the $x \to 0$ region, the distribution demonstrates problematic behavior due to the chosen renormalization scheme and LCDA models.

\begin{@twocolumnfalse}
    \begin{center}
        \begin{figure*}[htbp]
\centering
\begin{minipage}[t]{0.4\textwidth}
\centering
\includegraphics[width=7cm]{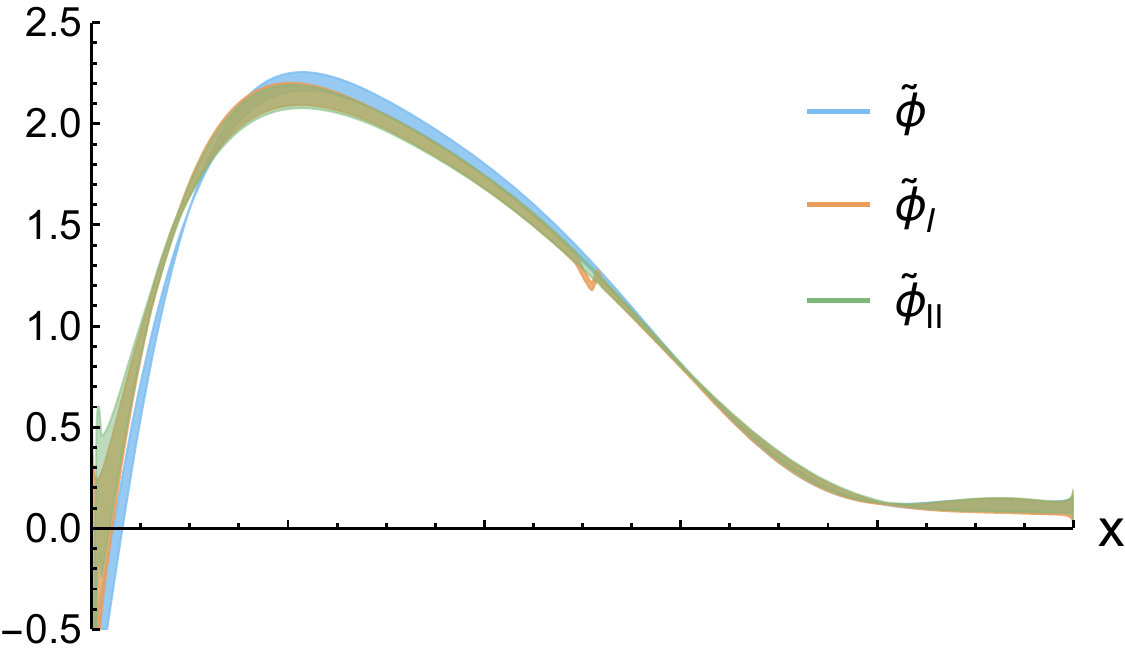}
\centerline{(a)}
\end{minipage}
\begin{minipage}[t]{0.5\textwidth}
\centering
\includegraphics[width=7cm]{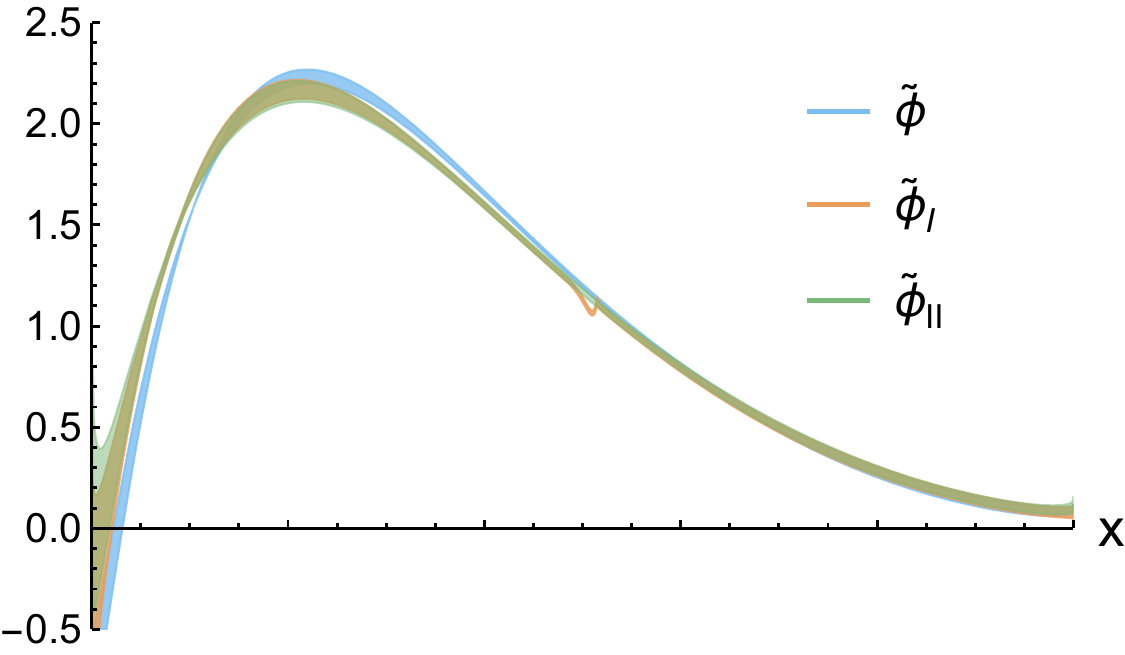}
\centerline{(b)}
\end{minipage}
\begin{minipage}[t]{0.4\textwidth}
\centering
\includegraphics[width=7cm]{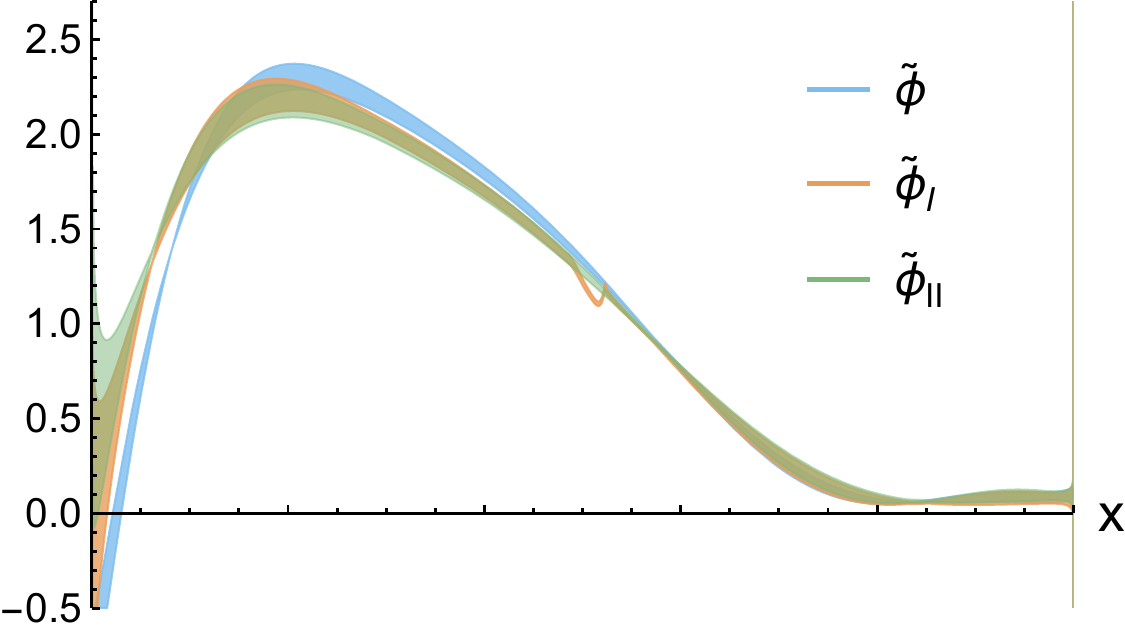}
\centerline{(c)}
\end{minipage}
\begin{minipage}[t]{0.5\textwidth}
\centering
\includegraphics[width=7cm]{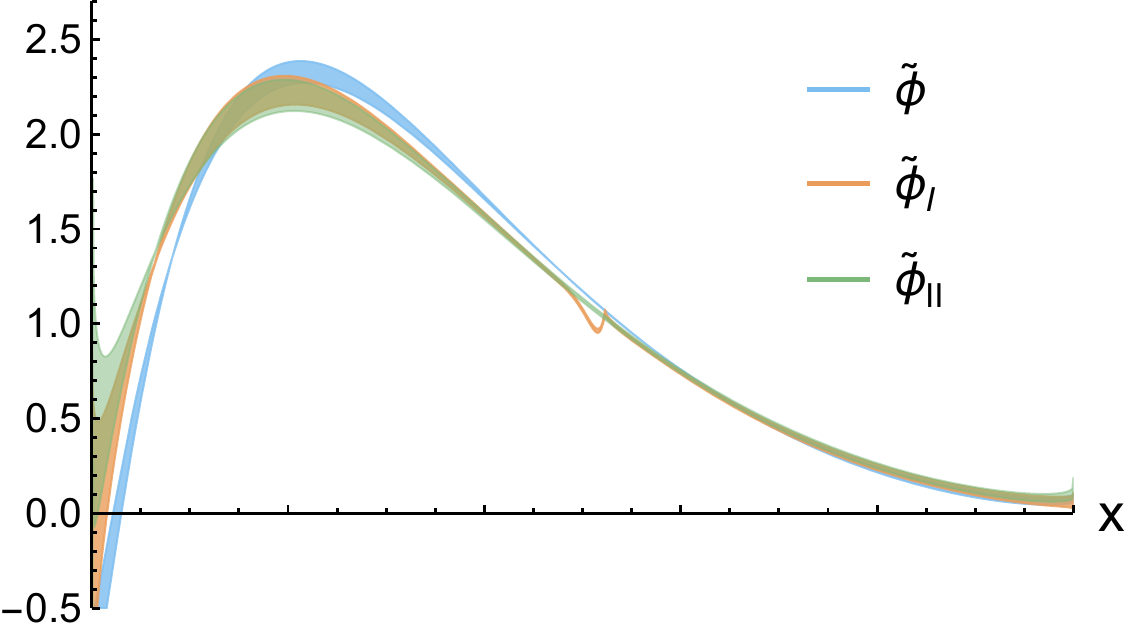}
\centerline{(d)}
\end{minipage}
\begin{minipage}[t]{0.4\textwidth}
\centering
\includegraphics[width=7cm]{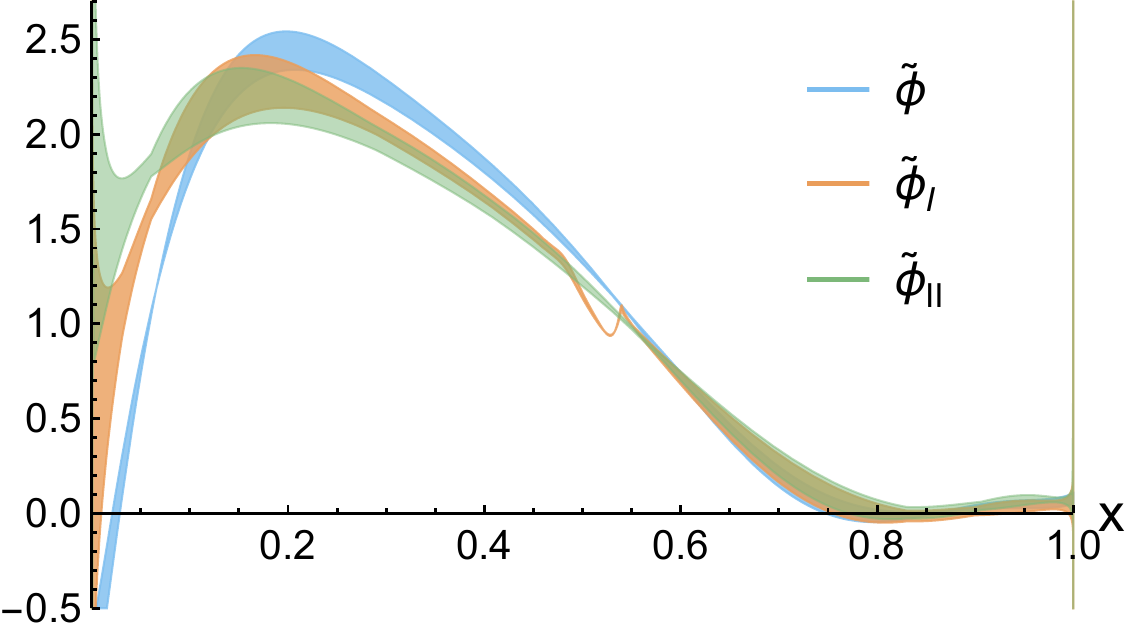}
\centerline{(e)}
\end{minipage}
\begin{minipage}[t]{0.5\textwidth}
\centering
\includegraphics[width=7cm]{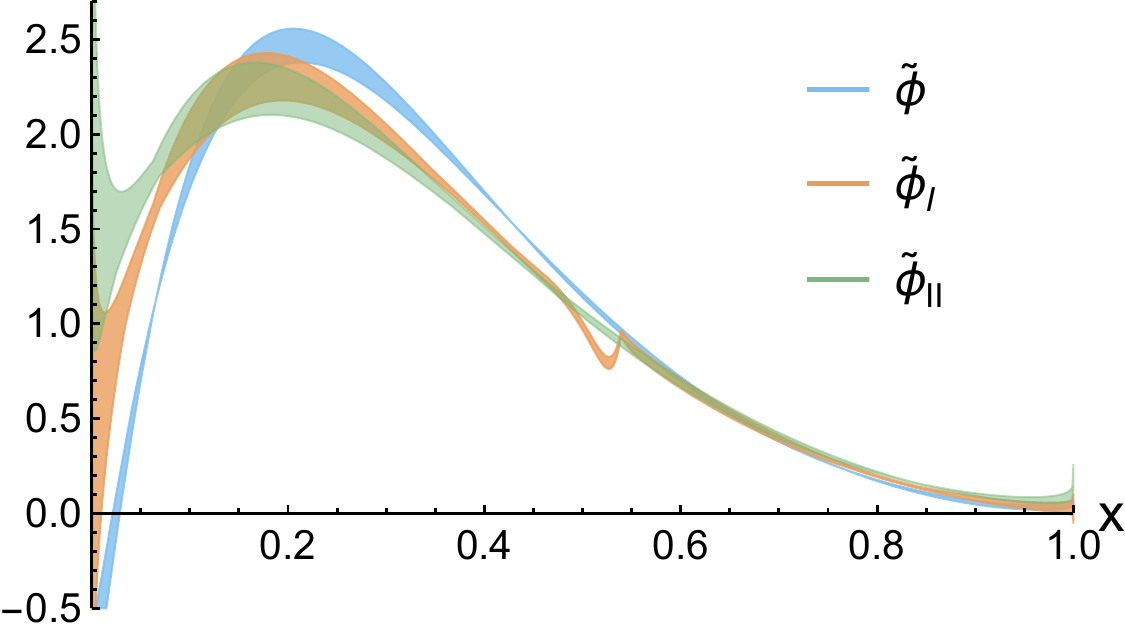}
\centerline{(f)}
\end{minipage}
\caption{
This figure compares renormalon-ambiguity-involved mass-corrected quasi-DAs with one-loop-corrected quasi-DAs.
The subfigures (a) and (b) display quasi-DAs at \( P^z = 5 \, \text{GeV} \), the subfigures (c) and (d) at \( P^z = 4 \, \text{GeV} \), and the subfigures (e) and (f) at \( P^z = 3 \, \text{GeV} \). 
The subfigures (a) (c) and (e) illustrate the impact of renormalon ambiguities on the COPE model, while the subfigures (b), (d) and (f) focus on the EXP model.
In all panels, \(\tilde{\phi}\) represents the one-loop corrected results. \(\tilde{\phi}_I\) denotes results mass-corrected through Eq.~(\ref{mc1}) incorporating renormalon ambiguities, and \(\tilde{\phi}_{II}\) shows results mass-corrected via Eq.~(\ref{mc2}) also including renormalon ambiguities.}
\label{amb-pic}
\end{figure*}
    \end{center}
\end{@twocolumnfalse}

\begin{@twocolumnfalse}
    \begin{center}
        \begin{figure*}[htbp]
\centering
\begin{minipage}[t]{0.4\textwidth}
\centering
\includegraphics[width=7cm]{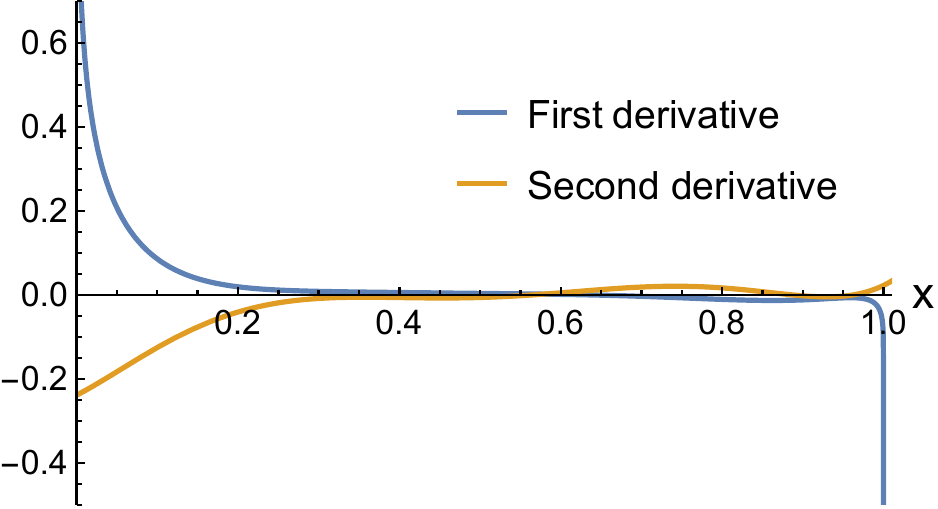}
\centerline{(a)}
\end{minipage}
\begin{minipage}[t]{0.5\textwidth}
\centering
\includegraphics[width=7cm]{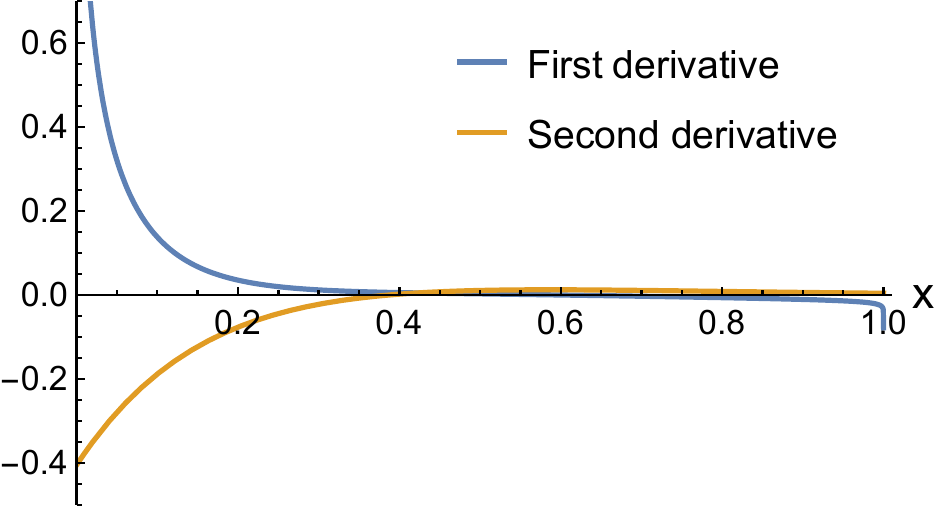}
\centerline{(b)}
\end{minipage}
\caption{These two picture show how the hybrid renormalization scheme modify the renormalon ambiguity. The blue lines show the contribution from first derivative and the orange lines show the contribution from the second derivative. The renormalon ambiguity to COPE model is shown in (a) and that to EXP model is shown in (b), both with $P^z=3~\mathrm{GeV}$ and $\mu=1.6~\mathrm{GeV}$. }
\label{amb2-pic}
\end{figure*}
    \end{center}
\end{@twocolumnfalse}

\begin{@twocolumnfalse}
    \begin{center}
        \begin{figure*}[htbp]
\centering
\begin{minipage}[t]{0.3\textwidth}
\centering
\includegraphics[width=5cm]{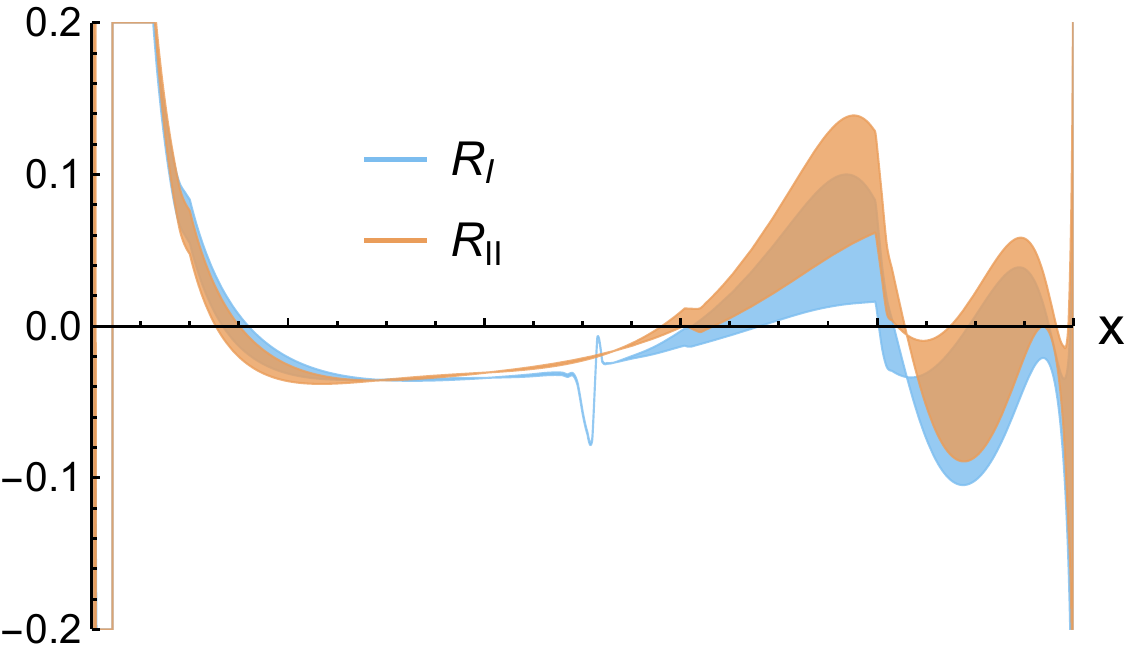}
\centerline{(a)}
\end{minipage}
\begin{minipage}[t]{0.3\textwidth}
\centering
\includegraphics[width=5cm]{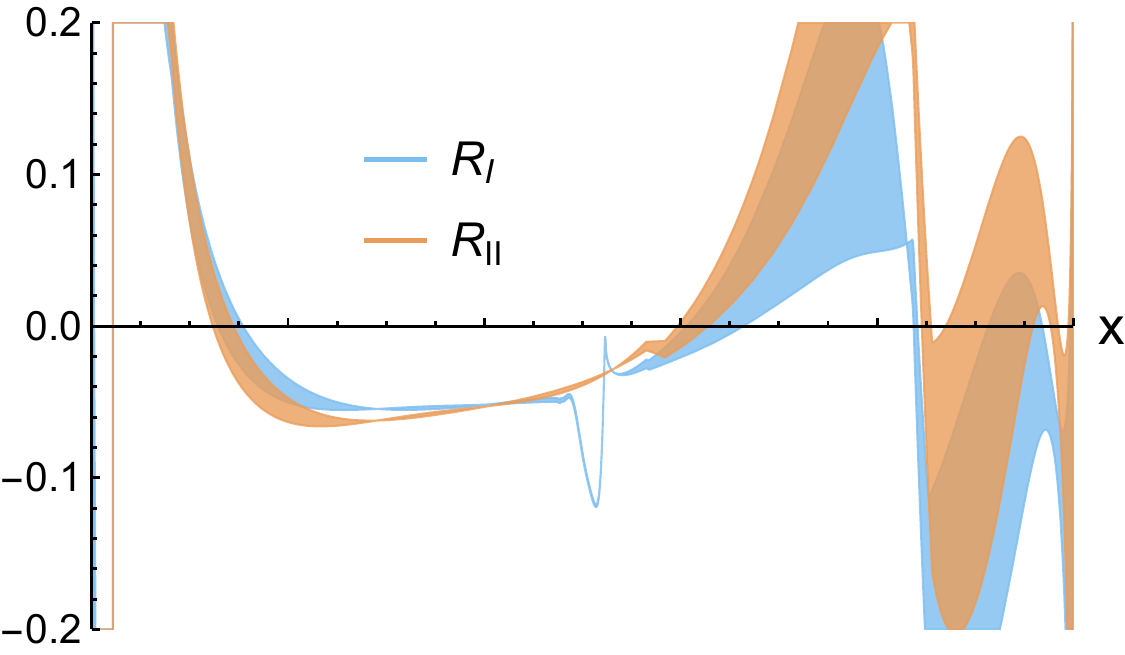}
\centerline{(b)}
\end{minipage}
\begin{minipage}[t]{0.3\textwidth}
\centering
\includegraphics[width=5cm]{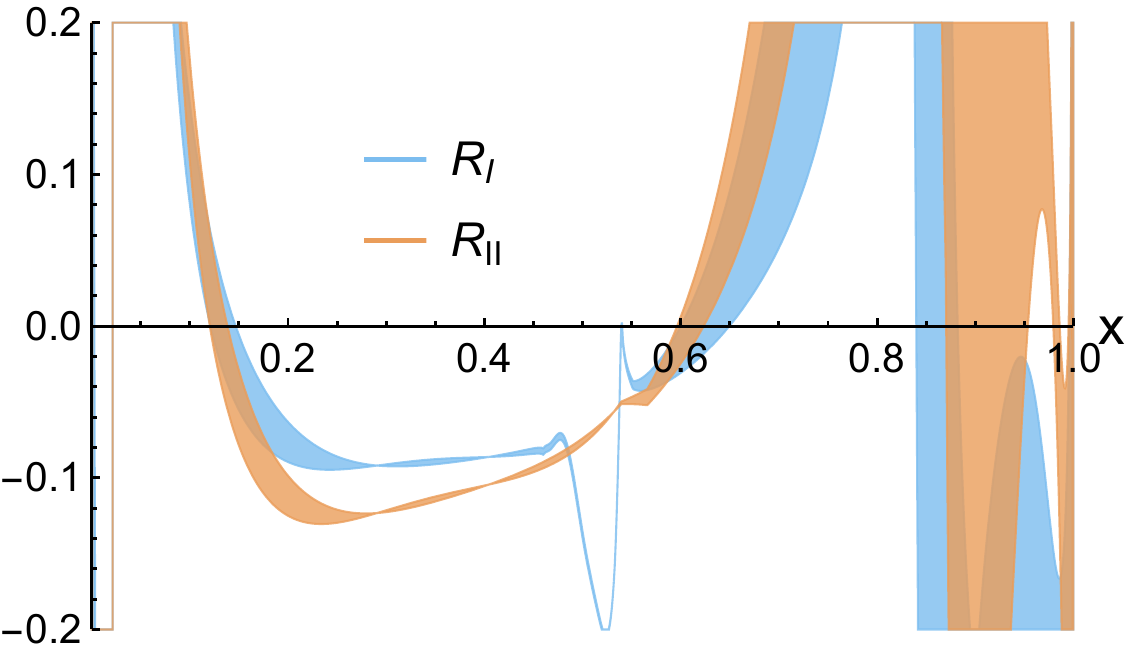}
\centerline{(c)}
\end{minipage}
\caption{Panel (a) corresponds to the ratio \( R_{I/II} \) at \( P^z = 5 \, \text{GeV} \), panel (b) to the ratio \( R_{I/II} \) at \( P^z = 4 \, \text{GeV} \), and panel (c) to the ratio \( R_{I/II} \) at \( P^z = 3 \, \text{GeV} \).}
\label{amb-pic-R}
\end{figure*}
    \end{center}
\end{@twocolumnfalse}

\subsection{$\Lambda_\text{QCD}^2/(P^z)^2$ Correction}
Finally, we address the \(\Lambda_\text{QCD}^2/(P^z)^2\) correction.
We adopt the formula presented in Eq.~(\ref{ambiguity}), which expands the error bands by \( \pm \mathcal{N} \frac{m_H^2}{(P^z)^2} \delta_R \phi(x, \mu) \). As mentioned in ~\cite{Braun:2024snf}, the renormalon ambiguity may be rather singular near the boundary of the Dokshitzer–Gribov–Lipatov–Altarelli–Parisi (DGLAP) and Efremov–Radyushkin–Brodsky–Lepage (ERBL) regions, depending the specific DA model chosen. For DAs, it happens to be $x\to 0$ and $x\to 1$, which necessitates special attention.
The renormalon model estimate of higher-twist corrections to different LCDA models and mass correction strategies at $P=\{3,4,5\}~\mathrm{GeV}$ are shown in Fig.~\ref{amb-pic}.
The $\Lambda_\text{QCD}^2/(P^z)^2$ correction is generally not as large as the mass correction in the moderate $x$ region, as expected from the naive power counting.
However, there is a significant enhancement in the former at the \(x \rightarrow 0\) endpoint region. 
As emphasized in Section III, the second derivative term comes from the hybrid renormalization scheme.
The contributions of the renormalons from both the first and second derivatives are depicted separately in Fig.~\ref{amb2-pic}, illustrating distinct effects.
One can observe that in both models, the hybrid renormalization scheme mitigates the ambiguity at the endpoint region $x\to 0$. Also, it magnifies the ambiguity at large $x$. 
However, the renormalon corrections at $x\to 1$ for the COPE model are larger than those for the EXP model.
The different asymptotic behaviors contained in the renormalon model can be found out by observing the shapes of the second derivative term of the two models in Fig.~\ref{amb2-pic}. We can observe that the contribution of second derivative is smoother comparing to that of COPE model, which we regard as a more physical behaviour.
        
Additionally, we define a ratio \( R_{I/II}=\displaystyle\frac{\tilde\phi_{I/II}-\tilde\phi}{\tilde\phi} \) to quantify the relative size of the corrections to the one-loop corrected results. The corresponding values of \( R_{I/II} \) for the specified analyses are depicted in Fig.~\ref{amb-pic-R}. 
{
The \(\Lambda_\text{QCD}^2/(P^z)^2\) correction depends  on the combination \(\Lambda_\text{QCD}/x P^z\) and \(\Lambda_\text{QCD}/((1-x) P^z)\). Therefore, a reliable region for the power correction \(\Lambda_\text{QCD}^2/(xP^z)^2\) and \(\Lambda_\text{QCD}^2/((1-x)P^z)^2\) is approximately \(x \in [0.2, 0.8]\) for our choice of parameters.} It can be observed that for the D-meson case, in the moderate \( x \) region $0.2<x<0.8$, the corrections from mass and power are typically smaller than 20\%. At the endpoint region,  the power corrections are sizable.

\section{Summary and Prospect}


In this paper, we have explored the impact of mass and ${\cal O}(\Lambda_\text{QCD}^2/(P^z)^2)$ corrections in the factorization between the heavy meson quasi-DAs and LCDAs within the LaMET framework.
For hadron mass corrections, we have employed two distinct methods. The first method relies on moments relations to provide mass corrections to all orders. The second method utilizes a leading-twist projector, a techniques employed for target mass corrections in DIS, to address leading-order mass corrections. The two methods yield consistent results when expanded to the same order.
For ${\cal O}(\Lambda_\text{QCD}^2/(P^z)^2)$ corrections, we have utilized the renormalon model. This model helps estimate the renormalon ambiguity through a single parameter, offering insights into the non-perturbative effects that influence our theoretical calculations.
Numerical calculations have been conducted based on these discussions. By employing two heavy meson LCDA models, we have investigated the impact of these corrections. As anticipated, the target mass corrections have a non-negligible effect, reinforcing the need for their inclusion in precision studies.

The findings from this work provide valuable guidance for future studies of heavy meson LCDAs, particularly in settings involving Lattice QCD. They also offer potential insights for phenomenological studies of heavy meson LCDAs.
However, there are still several issues that remain to be addressed in future analysis. 
Notably, the zero-momentum matrix element used in Eq.~(\ref{eq:hybrid_renorm}) for the hybrid renormalization scheme is from that calculated in the literature for a light meson, where the massive quark effects are omitted. 
The constituent mass corrections from the heavy quark need to be considered.
Furthermore, when performing the OPE for the LCDA case, the mass correction contributions from total derivative of the twist-2 term are significant and warrant further investigation to understand their full impact on the theory.

\section*{Acknowledgement}
We thank Xiangdong Ji, Yushan Su,  Qi-An Zhang and Jun Zeng for useful discussions. This work is supported in part by Natural Science Foundation of China under grant No. 12125503, 12335003, No. 12375080, No. 12061131006, 11975051, and by CUHK-Shenzhen under grant No. UDF01002851.

\appendix
\section{one-loop matching kernel}\label{app1}

The one-loop matching kernel in hybrid renormalization scheme reads~\cite{Ji:2020brr,Zhang:2023bxs,Baker:2024zcd}
\begin{align}
C\left(x, y, \frac{\mu}{P^z}\right)
=\delta(x-y)+C_B^{(1)}\left(x, y, \frac{\mu}{P^z}\right)-C_{C T}^{(1)}(x, y),
\end{align}
where
\begin{equation}
C_{C T}^{(1)}(x, y)=-\frac{3 \alpha_s C_F}{2 \pi^2}\left[\frac{\mathrm{Si}(z_s P^z(y-x))}{y-x}\right
]_{+(y)},
\end{equation}
and
\begin{align}\label{}
&C^{(1)}_B\left(\Gamma,x,y,\frac{P^z}{\mu}\right)
\\&
=\frac{\alpha_s C_F}{2\pi}\left\{
\begin{array}{lc}
\left[H_1(x,y)\right]_{+(y)}				& x<0<y\\
\left[H_2(x,y,P^z/\mu)\right]_{+(y)}		& 0<x<y\\
\left[H_2(1-x,1-y,P^z/\mu)\right]_{+(y)}	& y<x<1\\
\left[H_1(1-x,1-y)\right]_{+(y)}			& y<1<x
\end{array}\right. .
\end{align}
The sine integral function is defined as 
\begin{equation}
\mathrm{Si}(x)=\int_0^x\frac{\mathrm{sin}y}{y}dy.
\end{equation}
$H_1(x,y)$ and $H_2(x,y,\frac{P^z}{\mu})$ depend on the Dirac structure in the non-local operator. In this case we have
\begin{align}
H_1(x,y)&=\frac{1+x-y}{y-x}\frac{1-x}{1-y}\ln\frac{y-x}{1-x}
\\&
+\frac{1+y-x}{y-x}\frac{x}{y}\ln\frac{y-x}{-x},
\\
H_2\left(x,y,\frac{P^z}{\mu}\right)&=\frac{1+y-x}{y-x}\frac{x}{y}\ln\frac{4x(y-x)(P^z)^2}{\mu^2}
\\&
+\frac{1+x-y}{y-x}\left(\frac{1-x}{1-y}\ln\frac{y-x}{1-x}-\frac{x}{y}\right).
\end{align}
A plus function $h(x,y)_{+(y)}$is defined as 
\begin{equation}
    \begin{aligned}
        \int_0^1 dxh(x,y)_{+(y)}g(x)=\int_0^1 dx h(x,y)(g(x)-g(y)).
    \end{aligned}
\end{equation}
Utilizing this feature, the matching equation Eq.~(\ref{matching}) can be written as
\begin{equation}
 \begin{aligned}
 \Tilde{\phi}(x,P^z) &=\phi(x,\mu)\\
 &+\int_{-\infty}^{\infty} dy (\tilde C_B^{(1)}(x,y,\frac{\mu}{P^z})-\tilde C_{CT}^{(1)}(x,y))\phi(y,\mu) \\
 &-\int_{-\infty}^{\infty} dy(\tilde C_B^{(1)}(y,x,\frac{\mu}{P^z})-\tilde C_{CT}^{(1)}(y,x))\phi(x,\mu),
\end{aligned}   
\end{equation}
where $\tilde C_B^{(1)}(x,y,\frac{\mu}{P^z})$ and $\tilde C_{CT}^{(1)}(x,y)$ are defined as $C_B^{(1)}(x,y,\frac{\mu}{P^z})$ and $C_{CT}^{(1)}(x,y)$ without the plus part. These results have been applied in many analyses in LaMET~\cite{Zhang:2017bzy,Bali:2018spj,Xu:2018mpf,Zhang:2017zfe,Liu:2018tox,Wang:2019msf,Zhang:2020gaj,Hua:2020gnw,LatticeParton:2022zqc,Hu:2023bba,Deng:2023csv,Han:2023xbl,Han:2023hgy,Han:2024ucv}.


\begin{thebibliography}{}
\bibitem{Grozin:1996pq}
A.~G.~Grozin and M.~Neubert,
Phys. Rev. D \textbf{55}, 272-290 (1997)
doi:10.1103/PhysRevD.55.272
[arXiv:hep-ph/9607366 [hep-ph]].

\bibitem{Beneke:1999br}
M.~Beneke, G.~Buchalla, M.~Neubert and C.~T.~Sachrajda,
Phys. Rev. Lett. \textbf{83}, 1914-1917 (1999)
doi:10.1103/PhysRevLett.83.1914
[arXiv:hep-ph/9905312 [hep-ph]].

\bibitem{Beneke:2000ry}
M.~Beneke, G.~Buchalla, M.~Neubert and C.~T.~Sachrajda,
Nucl. Phys. B \textbf{591}, 313-418 (2000)
doi:10.1016/S0550-3213(00)00559-9
[arXiv:hep-ph/0006124 [hep-ph]].

\bibitem{Keum:2000wi}
Y.~Y.~Keum, H.~N.~Li and A.~I.~Sanda,
Phys. Rev. D \textbf{63}, 054008 (2001)
doi:10.1103/PhysRevD.63.054008
[arXiv:hep-ph/0004173 [hep-ph]].

\bibitem{Lu:2000em}
C.~D.~Lu, K.~Ukai and M.~Z.~Yang,
Phys. Rev. D \textbf{63}, 074009 (2001)
doi:10.1103/PhysRevD.63.074009
[arXiv:hep-ph/0004213 [hep-ph]].

\bibitem{Kawamura:2001jm}
H.~Kawamura, J.~Kodaira, C.~F.~Qiao and K.~Tanaka,
Phys. Lett. B \textbf{523}, 111 (2001)
[erratum: Phys. Lett. B \textbf{536}, 344-344 (2002)]
doi:10.1016/S0370-2693(01)01299-0
[arXiv:hep-ph/0109181 [hep-ph]].

\bibitem{Lange:2003ff}
B.~O.~Lange and M.~Neubert,
Phys. Rev. Lett. \textbf{91}, 102001 (2003)
doi:10.1103/PhysRevLett.91.102001
[arXiv:hep-ph/0303082 [hep-ph]].

\bibitem{Braun:2003wx}
V.~M.~Braun, D.~Y.~Ivanov and G.~P.~Korchemsky,
Phys. Rev. D \textbf{69}, 034014 (2004)
doi:10.1103/PhysRevD.69.034014
[arXiv:hep-ph/0309330 [hep-ph]].

\bibitem{Lee:2005gza}
S.~J.~Lee and M.~Neubert,
Phys. Rev. D \textbf{72}, 094028 (2005)
doi:10.1103/PhysRevD.72.094028
[arXiv:hep-ph/0509350 [hep-ph]].

\bibitem{Charng:2005fj}
Y.~Y.~Charng and H.~n.~Li,
Phys. Rev. D \textbf{72}, 014003 (2005)
doi:10.1103/PhysRevD.72.014003
[arXiv:hep-ph/0505045 [hep-ph]].

\bibitem{Kawamura:2008vq}
H.~Kawamura and K.~Tanaka,
Phys. Lett. B \textbf{673}, 201-207 (2009)
doi:10.1016/j.physletb.2009.02.028
[arXiv:0810.5628 [hep-ph]].

\bibitem{Bell:2013tfa}
G.~Bell, T.~Feldmann, Y.~M.~Wang and M.~W.~Y.~Yip,
JHEP \textbf{11}, 191 (2013)
doi:10.1007/JHEP11(2013)191
[arXiv:1308.6114 [hep-ph]].

\bibitem{Feldmann:2014ika}
T.~Feldmann, B.~O.~Lange and Y.~M.~Wang,
Phys. Rev. D \textbf{89}, no.11, 114001 (2014)
doi:10.1103/PhysRevD.89.114001
[arXiv:1404.1343 [hep-ph]].

\bibitem{Braun:2019wyx}
V.~M.~Braun, Y.~Ji and A.~N.~Manashov,
Phys. Rev. D \textbf{100}, no.1, 014023 (2019)
doi:10.3204/PUBDB-2019-02451
[arXiv:1905.04498 [hep-ph]].

\bibitem{Galda:2020epp}
A.~M.~Galda and M.~Neubert,
Phys. Rev. D \textbf{102}, 071501 (2020)
doi:10.1103/PhysRevD.102.071501
[arXiv:2006.05428 [hep-ph]].

\bibitem{Khodjamirian:2005ea}
A.~Khodjamirian, T.~Mannel and N.~Offen,
Phys. Lett. B \textbf{620}, 52-60 (2005)
doi:10.1016/j.physletb.2005.06.021
[arXiv:hep-ph/0504091 [hep-ph]].

\bibitem{Meissner:2013hya}
U.~G.~Mei\ss{}ner and W.~Wang,
Phys. Lett. B \textbf{730}, 336-341 (2014)
doi:10.1016/j.physletb.2014.02.009
[arXiv:1312.3087 [hep-ph]].

\bibitem{Gao:2019lta}
J.~Gao, C.~D.~L\"u, Y.~L.~Shen, Y.~M.~Wang and Y.~B.~Wei,
Phys. Rev. D \textbf{101}, no.7, 074035 (2020)
doi:10.1103/PhysRevD.101.074035
[arXiv:1907.11092 [hep-ph]].

\bibitem{Rahimi:2020zzo}
M.~Rahimi and M.~Wald,
Phys. Rev. D \textbf{104}, no.1, 016027 (2021)
doi:10.1103/PhysRevD.104.016027
[arXiv:2012.12165 [hep-ph]].

\bibitem{Khodjamirian:2023wol}
A.~Khodjamirian, B.~Meli\'c and Y.~M.~Wang,
Eur. Phys. J. ST \textbf{233}, no.2, 271-298 (2024)
doi:10.1140/epjs/s11734-023-01046-6
[arXiv:2311.08700 [hep-ph]].

\bibitem{Bell:2008er}
G.~Bell and T.~Feldmann,
JHEP \textbf{04}, 061 (2008)
doi:10.1088/1126-6708/2008/04/061
[arXiv:0802.2221 [hep-ph]].

\bibitem{Hwang:2010hw}
C.~W.~Hwang,
Phys. Rev. D \textbf{81}, 114024 (2010)
doi:10.1103/PhysRevD.81.114024
[arXiv:1003.0972 [hep-ph]].

\bibitem{Yang:2011ie}
M.~Z.~Yang,
Eur. Phys. J. C \textbf{72}, 1880 (2012)
doi:10.1140/epjc/s10052-012-1880-x
[arXiv:1104.3819 [hep-ph]].

\bibitem{Wu:2013lga}
X.~G.~Wu and T.~Huang,
Chin. Sci. Bull. \textbf{59}, 3801 (2014)
doi:10.1007/s11434-014-0335-1
[arXiv:1312.1455 [hep-ph]].

\bibitem{Sun:2016avp}
H.~K.~Sun and M.~Z.~Yang,
Phys. Rev. D \textbf{95}, no.11, 113001 (2017)
doi:10.1103/PhysRevD.95.113001
[arXiv:1609.08958 [hep-ph]].

\bibitem{Binosi:2018rht}
D.~Binosi, L.~Chang, M.~Ding, F.~Gao, J.~Papavassiliou and C.~D.~Roberts,
Phys. Lett. B \textbf{790}, 257-262 (2019)
doi:10.1016/j.physletb.2019.01.033
[arXiv:1812.05112 [nucl-th]].

\bibitem{Sun:2019xyw}
H.~K.~Sun and M.~Z.~Yang,
Phys. Rev. D \textbf{99}, no.9, 093002 (2019)
doi:10.1103/PhysRevD.99.093002
[arXiv:1903.04295 [hep-ph]].

\bibitem{Lan:2019img}
J.~Lan, C.~Mondal, M.~Li, Y.~Li, S.~Tang, X.~Zhao and J.~P.~Vary,
Phys. Rev. D \textbf{102}, no.1, 014020 (2020)
doi:10.1103/PhysRevD.102.014020
[arXiv:1911.11676 [nucl-th]].

\bibitem{Serna:2020txe}
F.~E.~Serna, R.~C.~da Silveira, J.~J.~Cobos-Mart\'\i{}nez, B.~El-Bennich and E.~Rojas,
Eur. Phys. J. C \textbf{80}, no.10, 955 (2020)
doi:10.1140/epjc/s10052-020-08517-3
[arXiv:2008.09619 [hep-ph]].

\bibitem{Shi:2021nvg}
C.~Shi, M.~Li, X.~Chen and W.~Jia,
Phys. Rev. D \textbf{104}, no.9, 094016 (2021)
doi:10.1103/PhysRevD.104.094016
[arXiv:2108.10625 [hep-ph]].

\bibitem{Feldmann:2022uok}
T.~Feldmann, P.~L\"ughausen and D.~van Dyk,
JHEP \textbf{10}, 162 (2022)
doi:10.1007/JHEP10(2022)162
[arXiv:2203.15679 [hep-ph]].

\bibitem{Serna:2022yfp}
F.~E.~Serna, R.~C.~da Silveira and B.~El-Bennich,
Phys. Rev. D \textbf{106}, no.9, L091504 (2022)
doi:10.1103/PhysRevD.106.L091504
[arXiv:2209.09278 [hep-ph]].

\bibitem{Almeida-Zamora:2023bqb}
B.~Almeida-Zamora, J.~J.~Cobos-Mart\'\i{}nez, A.~Bashir, K.~Raya, J.~Rodr\'\i{}guez-Quintero and J.~Segovia,
Phys. Rev. D \textbf{109}, no.1, 014016 (2024)
doi:10.1103/PhysRevD.109.014016
[arXiv:2309.17282 [hep-ph]].

\bibitem{Arifi:2024mff}
A.~J.~Arifi, L.~Happ, S.~Ohno and M.~Oka,
Phys. Rev. D \textbf{110}, no.1, 014020 (2024)
doi:10.1103/PhysRevD.110.014020
[arXiv:2401.07933 [hep-ph]].

\bibitem{Kawamura:2018gqz}
H.~Kawamura and K.~Tanaka,
PoS \textbf{RADCOR2017}, 076 (2018)
doi:10.22323/1.290.0076

\bibitem{Wang:2019msf}
W.~Wang, Y.~M.~Wang, J.~Xu and S.~Zhao,
Phys. Rev. D \textbf{102}, no.1, 011502 (2020)
doi:10.1103/PhysRevD.102.011502
[arXiv:1908.09933 [hep-ph]].

\bibitem{Zhao:2019elu}
S.~Zhao,
Phys. Rev. D \textbf{101}, no.7, 071503 (2020)
doi:10.1103/PhysRevD.101.071503
[arXiv:1910.03470 [hep-ph]].

\bibitem{Zhao:2020bsx}
S.~Zhao and A.~V.~Radyushkin,
Phys. Rev. D \textbf{103}, no.5, 054022 (2021)
doi:10.1103/PhysRevD.103.054022
[arXiv:2006.05663 [hep-ph]].

\bibitem{Xu:2022krn}
J.~Xu, X.~R.~Zhang and S.~Zhao,
Phys. Rev. D \textbf{106}, no.1, L011503 (2022)
doi:10.1103/PhysRevD.106.L011503
[arXiv:2202.13648 [hep-ph]].

\bibitem{Xu:2022guw}
J.~Xu and X.~R.~Zhang,
Phys. Rev. D \textbf{106}, no.11, 114019 (2022)
doi:10.1103/PhysRevD.106.114019
[arXiv:2209.10719 [hep-ph]].

\bibitem{Deng:2023csv}
Z.~F.~Deng, C.~Han, W.~Wang, J.~Zeng and J.~L.~Zhang,
JHEP \textbf{07}, 191 (2023)
doi:10.1007/JHEP07(2023)191
[arXiv:2304.09004 [hep-ph]].

\bibitem{Hu:2023bba}
S.~M.~Hu, W.~Wang, J.~Xu and S.~Zhao,
Phys. Rev. D \textbf{109}, no.3, 034001 (2024)
doi:10.1103/PhysRevD.109.034001
[arXiv:2308.13977 [hep-ph]].

\bibitem{Han:2023xbl}
C.~Han, Y.~Su, W.~Wang and J.~L.~Zhang,
JHEP \textbf{12}, 044 (2023)
doi:10.1007/JHEP12(2023)044
[arXiv:2308.16793 [hep-ph]].

\bibitem{Han:2023hgy}
C.~Han and J.~Zhang,
Phys. Rev. D \textbf{109}, no.1, 014034 (2024)
doi:10.1103/PhysRevD.109.014034
[arXiv:2311.02669 [hep-ph]].

\bibitem{Hu:2024ebp}
S.~M.~Hu, J.~Xu and S.~Zhao,
Eur. Phys. J. C \textbf{84}, no.5, 502 (2024)
doi:10.1140/epjc/s10052-024-12672-2
[arXiv:2401.04291 [hep-ph]].

\bibitem{Han:2024ucv}
C.~Han, W.~Wang, J.~Zeng and J.~L.~Zhang,
JHEP \textbf{07}, 019 (2024)
doi:10.1007/JHEP07(2024)019
[arXiv:2404.04855 [hep-ph]].

\bibitem{Han:2024min}
X.~Y.~Han, J.~Hua, X.~Ji, C.~D.~L\"u, W.~Wang, J.~Xu, Q.~A.~Zhang and S.~Zhao,
[arXiv:2403.17492 [hep-ph]].

\bibitem{Ji:2013dva}
X.~Ji,
Phys. Rev. Lett. \textbf{110}, 262002 (2013)
doi:10.1103/PhysRevLett.110.262002
[arXiv:1305.1539 [hep-ph]].

\bibitem{Ji:2014gla}
X.~Ji,
Sci. China Phys. Mech. Astron. \textbf{57}, 1407-1412 (2014)
doi:10.1007/s11433-014-5492-3
[arXiv:1404.6680 [hep-ph]].

\bibitem{Ji:2020ect}
X.~Ji, Y.~S.~Liu, Y.~Liu, J.~H.~Zhang and Y.~Zhao,
Rev. Mod. Phys. \textbf{93}, no.3, 035005 (2021)
doi:10.1103/RevModPhys.93.035005
[arXiv:2004.03543 [hep-ph]].

\bibitem{Ishaq:2019dst}
S.~Ishaq, Y.~Jia, X.~Xiong and D.~S.~Yang,
Phys. Rev. Lett. \textbf{125}, no.13, 132001 (2020)
doi:10.1103/PhysRevLett.125.132001
[arXiv:1905.06930 [hep-ph]].

\bibitem{Beneke:2023nmj}
M.~Beneke, G.~Finauri, K.~K.~Vos and Y.~Wei,
JHEP \textbf{09}, 066 (2023)
doi:10.1007/JHEP09(2023)066
[arXiv:2305.06401 [hep-ph]].

\bibitem{Han:2024yun}
X.~Y.~Han, J.~Hua, X.~Ji, C.~D.~L\"u, A.~Sch\"afer, Y.~Su, W.~Wang, J.~Xu, Y.~Yang and J.~H.~Zhang, \textit{et al.}
[arXiv:2410.18654 [hep-lat]].

\bibitem{Deng:2024dkd}
Z.~F.~Deng, W.~Wang, Y.~B.~Wei and J.~Zeng,
[arXiv:2409.00632 [hep-ph]].

\bibitem{Zhang:2017bzy}
J.~H.~Zhang, J.~W.~Chen, X.~Ji, L.~Jin and H.~W.~Lin,
Phys. Rev. D \textbf{95}, no.9, 094514 (2017)
doi:10.1103/PhysRevD.95.094514
[arXiv:1702.00008 [hep-lat]].

\bibitem{Chen:2016utp}
J.~W.~Chen, S.~D.~Cohen, X.~Ji, H.~W.~Lin and J.~H.~Zhang,
Nucl. Phys. B \textbf{911}, 246-273 (2016)
doi:10.1016/j.nuclphysb.2016.07.033
[arXiv:1603.06664 [hep-ph]].

\bibitem{Beneke:1994sw}
M.~Beneke and V.~M.~Braun,
Nucl. Phys. B \textbf{426}, 301-343 (1994)
doi:10.1016/0550-3213(94)90314-X
[arXiv:hep-ph/9402364 [hep-ph]].

\bibitem{Braun:2018brg}
V.~M.~Braun, A.~Vladimirov and J.~H.~Zhang,
Phys. Rev. D \textbf{99}, no.1, 014013 (2019)
doi:10.1103/PhysRevD.99.014013
[arXiv:1810.00048 [hep-ph]].

\bibitem{Braun:2024snf}
V.~M.~Braun, M.~Koller and J.~Schoenleber,
Phys. Rev. D \textbf{109}, no.7, 074510 (2024)
doi:10.1103/PhysRevD.109.074510
[arXiv:2401.08012 [hep-ph]].

\bibitem{Ji:2020brr}
X.~Ji, Y.~Liu, A.~Sch\"afer, W.~Wang, Y.~B.~Yang, J.~H.~Zhang and Y.~Zhao,
Nucl. Phys. B \textbf{964}, 115311 (2021)
doi:10.1016/j.nuclphysb.2021.115311
[arXiv:2008.03886 [hep-ph]].

\bibitem{Zhang:2023bxs}
R.~Zhang, J.~Holligan, X.~Ji and Y.~Su,
Phys. Lett. B \textbf{844}, 138081 (2023)
doi:10.1016/j.physletb.2023.138081
[arXiv:2305.05212 [hep-lat]].

\bibitem{Baker:2024zcd}
E.~Baker, D.~Bollweg, P.~Boyle, I.~Clo\"et, X.~Gao, S.~Mukherjee, P.~Petreczky, R.~Zhang and Y.~Zhao,
JHEP \textbf{07}, 211 (2024)
doi:10.1007/JHEP07(2024)211
[arXiv:2405.20120 [hep-lat]].

\bibitem{Ma:2017pxb}
Y.~Q.~Ma and J.~W.~Qiu,
Phys. Rev. Lett. \textbf{120}, no.2, 022003 (2018)
doi:10.1103/PhysRevLett.120.022003
[arXiv:1709.03018 [hep-ph]].

\bibitem{Anikin:1978tj}
S.~A.~Anikin and O.~I.~Zavyalov,
Annals Phys. \textbf{116}, 135-166 (1978)
doi:10.1016/0003-4916(78)90007-6

\bibitem{Anikin:1979kq}
S.~A.~Anikin, O.~I.~Zavyalov and N.~I.~Karchev,
Theor. Math. Phys. \textbf{38}, 193-202 (1979)
doi:10.1007/BF01018535

\bibitem{Balitsky:1987bk}
I.~I.~Balitsky and V.~M.~Braun,
Nucl. Phys. B \textbf{311}, 541-584 (1989)
doi:10.1016/0550-3213(89)90168-5

\bibitem{Muller:1994ses}
D.~M\"uller, D.~Robaschik, B.~Geyer, F.~M.~Dittes and J.~Ho\v{r}ej\v{s}i,
Fortsch. Phys. \textbf{42}, 101-141 (1994)
doi:10.1002/prop.2190420202
[arXiv:hep-ph/9812448 [hep-ph]].

\bibitem{Balitsky:1990ck}
I.~I.~Balitsky and V.~M.~Braun,
Nucl. Phys. B \textbf{361}, 93-140 (1991)
doi:10.1016/0550-3213(91)90618-8

\bibitem{Geyer:1999uq}
B.~Geyer, M.~Lazar and D.~Robaschik,
Nucl. Phys. B \textbf{559}, 339-377 (1999)
doi:10.1016/S0550-3213(99)00334-X
[arXiv:hep-th/9901090 [hep-th]].

\bibitem{Ball:1998kk}
P.~Ball and V.~M.~Braun,
Phys. Rev. D \textbf{58}, 094016 (1998)
doi:10.1103/PhysRevD.58.094016
[arXiv:hep-ph/9805422 [hep-ph]].

\bibitem{Braun:2011zr}
V.~M.~Braun and A.~N.~Manashov,
Phys. Rev. Lett. \textbf{107}, 202001 (2011)
doi:10.1103/PhysRevLett.107.202001
[arXiv:1108.2394 [hep-ph]].

\bibitem{Braun:2011dg}
V.~M.~Braun and A.~N.~Manashov,
JHEP \textbf{01}, 085 (2012)
doi:10.1007/JHEP01(2012)085
[arXiv:1111.6765 [hep-ph]].

\bibitem{Bali:2018spj}
G.~S.~Bali, V.~M.~Braun, B.~Gl\"a\ss{}le, M.~G\"ockeler, M.~Gruber, F.~Hutzler, P.~Korcyl, A.~Sch\"afer, P.~Wein and J.~H.~Zhang,
Phys. Rev. D \textbf{98}, no.9, 094507 (2018)
doi:10.1103/PhysRevD.98.094507
[arXiv:1807.06671 [hep-lat]].

\bibitem{Xu:2018mpf}
J.~Xu, Q.~A.~Zhang and S.~Zhao,
Phys. Rev. D \textbf{97}, no.11, 114026 (2018)
doi:10.1103/PhysRevD.97.114026
[arXiv:1804.01042 [hep-ph]].

\bibitem{Zhang:2017zfe}
J.~H.~Zhang \textit{et al.} [LP3],
Nucl. Phys. B \textbf{939}, 429-446 (2019)
doi:10.1016/j.nuclphysb.2018.12.020
[arXiv:1712.10025 [hep-ph]].

\bibitem{Liu:2018tox}
Y.~S.~Liu, W.~Wang, J.~Xu, Q.~A.~Zhang, S.~Zhao and Y.~Zhao,
Phys. Rev. D \textbf{99}, no.9, 094036 (2019)
doi:10.1103/PhysRevD.99.094036
[arXiv:1810.10879 [hep-ph]].

\bibitem{Zhang:2020gaj}
R.~Zhang, C.~Honkala, H.~W.~Lin and J.~W.~Chen,
Phys. Rev. D \textbf{102}, no.9, 094519 (2020)
doi:10.1103/PhysRevD.102.094519
[arXiv:2005.13955 [hep-lat]].

\bibitem{Hua:2020gnw}
J.~Hua \textit{et al.} [Lattice Parton],
Phys. Rev. Lett. \textbf{127}, no.6, 062002 (2021)
doi:10.1103/PhysRevLett.127.062002
[arXiv:2011.09788 [hep-lat]].

\bibitem{LatticeParton:2022zqc}
J.~Hua \textit{et al.} [Lattice Parton],
Phys. Rev. Lett. \textbf{129}, no.13, 132001 (2022)
doi:10.1103/PhysRevLett.129.132001
[arXiv:2201.09173 [hep-lat]].
\end{thebibliography}
\end{document}